\documentclass[a4paper,11pt]{article}

\usepackage{jcappub} 

\usepackage[T1]{fontenc} 

\usepackage[utf8]{inputenc} 		
\usepackage[ngerman,english]{babel} 
\usepackage[T1]{fontenc}    		
\usepackage{outlines}
\usepackage{amsmath,amssymb}
\usepackage{graphicx}
\usepackage{slashed}
\usepackage[small, bf, flushleft]{caption}
\usepackage{subcaption}
\usepackage{physics}
\usepackage{booktabs}
\usepackage{multirow}
\usepackage[dvipsnames]{xcolor}

\usepackage{hhline}
\usepackage{hyperref}
\hypersetup{
    colorlinks,
    citecolor=black,
    filecolor=black,
    linkcolor=blue,
    urlcolor=blue
}

\newcommand{\D}{\mathrm{d}}
\newcommand{\LCDM}{$\Lambda$CDM}

\newcommand{\eV}{\,\mathrm{eV}}
\newcommand{\GeV}{\,\mathrm{GeV}}

\newcommand{\Msol}{M_{\odot}}
\newcommand{\Mpc}{\,\text{Mpc}}
\newcommand{\kmsmpc}{\,\frac{\mathrm{km}/\mathrm{s}}{\Mpc}}

\title{\boldmath Probing alternative cosmologies through the inverse distance ladder}

\author[b]{Manfred Lindner}
\author[a]{Kevin Max}
\author[b]{Moritz Platscher}
\author[b]{Jonas Rezacek}

\affiliation[a]{Scuola Normale Superiore and INFN Pisa,\\
Piazza dei Cavalieri, 7 - 56126 Pisa, Italy}
\affiliation[b]{Max-Planck-Institut f\"ur Kernphysik, \\
Saupfercheckweg 1, 69117 Heidelberg, Germany}

\emailAdd{lindner@mpi-hd.mpg.de}
\emailAdd{kevin.max@sns.it}
\emailAdd{moritz.platscher@mpi-hd.mpg.de}
\emailAdd{jonas.rezacek@mpi-hd.mpg.de}

\abstract{
We study the implications of a combined analysis of cosmic standard candles and standard rulers on the viability of cosmological models beyond the cosmological concordance model \LCDM. To this end, we employ well-established data in the form of the joint light-curve analysis supernova compilation, baryon acoustic oscillations, and cosmic microwave background data on the one hand, and a recently proposed set of quasars as objects of known brightness on the other hand.
The advantage of including the latter is that they extend the local distance measures to redshifts which have previously been out of reach and we investigate how this allows one to test cosmologies beyond \LCDM. 
While there exist various studies on parametric extensions of \LCDM, we present here a comparative study of both parametric and fundamental extensions of the standard cosmology. 
In order to keep the scope of this manuscript contained, we focus on two particular modifications: One is the consistent theory of two interacting spin-2 objects, so-called bigravity, and the other conformal gravity, a theory of gravity that has no knowledge of fundamental length scales. 
The former of the two constitutes a veritable extension of General Relativity, given that it adds to the metric tensor of gravity a second dynamical tensor field. The resulting dynamics have been proposed as a self-accelerating cosmology.
Conformal gravity on the other hand is a much more drastic change of the underlying gravitational theory. Its ignorance towards fundamental length scales offers a completely different approach to the origin of late time acceleration. 
In this sense, both models offer --~in one way or another~-- an explanation for the dark energy problem. 
We perform a combined cosmological fit which provides strong constraints on these extensions.
We also briefly comment on the implications of the long-standing $H_0$-tension.
}

\begin{document}
\maketitle
\flushbottom

\section{Introduction}

For many years, cosmology was the driving force that sparked many ideas addressing the dark matter (DM) and dark energy (DE) problems in high energy physics, as well as the need for explaining baryogenesis on a microscopic scale. In recent years, however, it has become a precision discipline of fundamental physics itself.
Much like the recent advances in experimental particle physics have allowed to differentiate between microscopic models, cosmological analyses can now identify and rule out certain cosmological models. 
Nevertheless, existing studies of cosmological data often study parametrised modifications of the concordance cosmology of a flat universe with cold DM and a static cosmological constant (CC) $\Lambda$, henceforth \LCDM. This follows the same spirit as effective field theories in high energy physics, which are used to study a larger class of models that give rise to the same or similar low energy phenomena. 
While this is an important cornerstone in the survey of physics beyond the standard model (SM), it is indispensable to try and disentangle these effective descriptions at the fundamental level, as they may originate from very different fundamental principles with potentially grave impact on the particle physics sector.

Given the prevailing absence of \emph{particle} physics beyond the SM, it is a timely question to ask if and how well \emph{gravitational} physics beyond the SM can address these open questions, and possibly deliver answers to the open questions of the Universe.
In the present manuscript, we wish to help bridge this gap between parametrised physics beyond standard cosmology and fundamental models, building upon some previous analyses that have studied such modifications in isolation, such as $f(R)$~\cite{Leanizbarrutia:2017xyd,Nunes:2016drj,Odintsov:2017qif,Hagstotz:2018onp,Chen:2019uci} and $f(T)$  gravity~\cite{Bellini:2015xja,Nunes:2018xbm,Akarsu:2019ygx}, Brans-Dicke gravity~\cite{Perez:2018qgw,Sola:2019jek}, Galileons and Horndeski gravity~\cite{Kreisch:2017uet,Renk:2017rzu,Noller:2018wyv,Leloup:2019fas}, Quintessence~\cite{Yang:2018xah}, some combinations of these~\cite{Raveri:2019mxg}, and even non-local gravity~\cite{Dirian:2016puz,Belgacem:2017cqo,Amendola:2019fhc} to name only a few recent examples. 
Our aim is to deliver a blueprint to study fundamental modifications of the gravitational sector, and choose among the theories the one that best explains the data. In this we rely on statistical methods and available cosmological and astrophysical data sets, which we apply to \LCDM~and some of its parametrised extensions, as well as two fundamental cosmologies. 

The data sets under consideration include the well-known Supernovae (SNe) type-I data that first hinted at the accelerated expansion of the Universe at late times and thus manifested the dark energy problem in the late 1990s~\cite{Riess:1998cb,Perlmutter:1998np}. We also include data from quasar surveys which have only recently been shown to serve as standard candles, however, tracing the expansion of the Universe to much larger redshift than SNe~\cite{Risaliti:2015zla,Risaliti:2018reu}. Very recently, these were used to obtain new constraints on the standard cosmological model in Ref.~\cite{Khadka:2019njj}. Furthermore, we use data from galaxy and Lyman-$\alpha$ (Ly$\alpha$) surveys that extract from the clustering of matter the scale of baryon acoustic oscillations (BAO) in the early Universe. Finally, this sample is enhanced by the measurement of the acoustic scale from the CMB by the Planck Collaboration~\cite{Aghanim:2018eyx}. In using these data sets we employ the so-called inverse distance ladder method, meaning that we calibrate the (unknown) absolute magnitude of the standard candles via cosmic standard rulers, see e.g.~Refs.~\cite{Aubourg:2014yra,Feeney:2018mkj,Macaulay:2018fxi} for similar studies.

As for the models, we compare parametric extensions of the concordance model to models which are constructed from a symmetry or particle content point-of-view; i.e.~we study bigravitational cosmologies, and moreover conformal gravity (CG) cosmology. We have chosen these two models because they are sufficiently complementary  in their construction and phenomenology and will allow us to exemplify the usefulness of cosmological data applied to microscopic models. At the same time, both models are apt to address the problem associated with late-time acceleration, i.e.~the identity of DE. 

Bigravity is a modified version of the de Rham--Gabadadze--Tolley  (dRGT) theory of massive gravity~\cite{deRham:2010ik,deRham:2010kj,deRham:2011rn,Hassan:2011hr,Hassan:2011vm,Hassan:2011tf,Mirbabayi:2011aa,Comelli:2012vz,Deffayet:2012nr,Deffayet:2012zc}, in which the auxiliary tensor is dynamical~\cite{Hassan:2011zd,Hassan:2011ea}. This modification of the gravitational sector has profound consequences for the dynamics of the Universe: First, it introduces what may be considered either a tensor field that is closer in spirit to the matter fields that couple minimally to the physical metric, or indeed a second, \emph{hidden} metric which could serve as a metric for a dark sector, constructed in such a way that the DM in this sector communicates with the SM exclusively via gravity. Second, the massive spin-2 state, present in addition to the massless spin-2 graviton, dynamically sets the scale of late-time acceleration. If the mass is of order of the Hubble rate today, not only can this parametrise the late-time acceleration, but also stabilize it against radiative corrections. In this model, the CC problem is reduced to the question why the vacuum energy of the matter Lagrangian does not yield large corrections to the CC.

CG, on the other hand, is an attempt to abolish all scales in nature on a fundamental level by means of a symmetry principle: A conformally invariant action cannot contain any dimensionful parameters or couplings in four dimensions. Thus, the issue of small mass scales in nature (often termed a hierarchy problem with respect to the large Planck mass in General Relativity (GR)) is diverted to the problem of generating such scales dynamically, for which several mechanisms are well-known. In such conformal models of gravity, like e.g.~\cite{Zee:1978wi, Adler:1982ri}, it is often argued that GR is recovered as an effective theory at low energies, so no significant deviations from the standard \LCDM\ scenario are to be expected. In other realizations, e.g. Ref.~\cite{Mannheim:2005bfa}, different predictions for cosmic expansion arise than in the \LCDM\ model. At the same time these predictions are well testable with the aforementioned data sets and this therefore constitutes an interesting model to study in this work.
The absence of scales on a fundamental level has the amusing consequence that the theory and its cosmological solution are sensitive to systems of very different length scales, such as the scale of BAO on the far end of scales, but also to galaxy dynamics. In fact, it has been prominently argued in the past that CG can address the missing mass problem --~at least at the galactic level manifesting as the rise in rotation curves~\cite{Mannheim:1988dj}.

Our work extends the existing literature in several directions. First, it is the first time that the recent distance measurements of quasars are used in conjunction with SN data, BAO data and $H(z)$ measurements. Moreover, we do not only consider parametrised extensions of \LCDM, but take a close look at two microscopic generalizations of GR that give rise to modified cosmologies. One of them, CG, is a theory devoid of any fundamental length scale. We feel that the literature lacks a statistically sound and, as far as cosmology is concerned, comprehensive study of this model exploiting available cosmological data. The strong claims about CG, i.e.~being able to reconcile late time acceleration and the dark matter puzzle, call for such a comprehensive survey.

On the other hand we have bigravity, which is studied within the only known, fully analytical regime. Previous work performing a statistical analysis of various cosmological observations within bigravity can be found in \cite{Akrami:2012vf}; our analysis improves on this by assuming a less restrictive parameter set (non-zero curvature and radiation density). We furthermore present a completely new approach to solving the non-linear dynamical equations and identify the relevant physical branches. This combined effort allows us to draw some important conclusions on the model parameters and on the models themselves.

Note that in performing this cosmological fit, we do not address the current $H_0$-tension, i.e.~the discrepancy at the level of $4.4\,\sigma$ between the measurement of $H_0$ inferred from the Planck CMB ($H_0 = 67.4 \pm 0.5\,\frac{\text{km}}{\text{s}\Mpc}$~\cite{Aghanim:2018eyx}) and Supernova ($H_0 = 73.24 \pm 1.74\,\frac{\text{km}}{\text{s}\Mpc}$~\cite{Riess:2016jrr}) observations at $z=\order{0.01 - 0.1}$. SNe on their own do not constrain $H_0$; the tension arises only when the data is calibrated, for example using nearby Cepheids in the same host galaxy~\cite{Riess:2019cxk} (see also App.~\ref{subsec:SN_data}); a high value of $H_0$ in agreement with this is also found via time delays in gravitational lensing~\cite{Wong:2019kwg}. Reproducing the calibrated SN + Cepheid measurement of $H_0$ is beyond the scope of this work and a solution of the tension is not hinted at by the best fits of our modified cosmologies. This is in agreement with a recent study~\cite{Dhawan:2020xmp}, which finds that the locally measured value of $H_0$ remains largely unaffected by the choice of exotic background cosmologies; in particular, a variant of bimetric gravity is tested, and no alleviation of the $H_0$-tension is found.

We also point out the recent discussion which revolves around a possible discrepancy between \LCDM~and the high-redshift quasar data set. The authors of \cite{Risaliti:2018reu} fit a concordance model to the SN + quasar data up to $z<1.4$. They then extrapolate the model to the whole redshift region of the quasar data and, perhaps unsurprisingly, find that model and data are in $\approx 4\,\sigma$-tension at high-$z$. Finally, the authors perform an expansion of the Hubble function at low-$z$ and fit the Taylor coefficients to the data (`\textit{cosmographic expansion}'), and reach the conclusion that this can alleviate the tension. Several other works have appeared in the meantime, re-analysing these claims. In particular, if the quasar data is calibrated by the distance modulus of the SNe, as in the original study, the tension is reproduced~\cite{Yang:2019vgk,Velten:2019vwo}.
However, if the fit is performed including all parameters required to calculate the distance modulus from the raw quasar data (i.e.~including nuisance parameters), no such tension arises.
In this work, we adopt the latter viewpoint. As we will see, comparing the SN and joint SN + quasar fits then does not lead to a discrepancy for \LCDM.

Before moving on, we point out further data sets which we do not consider in our analysis:

\noindent While we incorporate the acoustic scale measurement from CMB, a full analysis including the matter fluctuation amplitude $\sigma_8$ is not performed. Planck has measured $\sigma_8$ to very high accuracy, which can be used to improve constraints of scenarios beyond SM plus GR, see for example~\cite{DiValentino:2015wba}. However a moderate tension arises when compared to observations of large scale structures, see DES lensing results \cite{Aghanim:2018eyx,Abbott:2017wau} for the current status. The tension has called for various theoretical models which modify the late-time universe compared to \LCDM; a feature which may be provided by bigravity as well as CG, and in the case of $\omega$\LCDM~has shown to slightly alleviate the tension~\cite{Lambiase:2018ows}. However a full analysis of the CMB spectrum is beyond the scope of this work. {In particular, the analysis of perturbations to the cosmological background solution within bigravity has proven a difficult task, as linear perturbations show unstable scalar and/or tensor modes~\cite{Konnig:2015lfa}. See Sec.~\ref{sec:bigra_cosmology} for a discussion of the branches and stability requirements in bigravity.}

Another promising avenue is that of velocity-induced acoustic oscillations (VAO) due to relative velocity between baryons and DM. Using data from the upcoming HERA interferometer, the Hubble function could be probed up to redshift $z=15 - 20$~\cite{Munoz:2019fkt}; however, data is not yet available. While the recent discovery of gravitational waves has opened a new window on the cosmological history of the universe, current observations are not yet precise enough to improve constraints on cosmological parameters. We also do not make use of extragalactic background light data measured by Fermi-LAT \cite{Dominguez:2019jqc} or gamma-ray bursts, which can be observed up to redshift $z \approx 6$~\cite{Wei:2013xx}. In both cases, inclusion of the data sets does not promise to increase the precision, nor does it extend the redshift-range of the test. 

This paper is structured as follows: in Sec.~\ref{sec:cosmological_models}, we review the basics of the models we analyse, and construct the Hubble function which enters the cosmological fit. In the case of CG, this entails a discussion whether the galactic rotation curves may be explained without DM. In Sec.~\ref{sec:results}, we discuss the results of the cosmological fits, and present the Hubble diagrams as well as the posterior probability distributions for the relevant parameters. We furthermore compare the competing theories and explain which model is favoured by statistical evidence. We draw our conclusions in Sec.~\ref{sec:conclusions}. In App.~\ref{app:data}, we discuss the different data sets we include and lay out the details of the Bayesian analysis as well as any other physical constraints.

\section{Cosmological models}
\label{sec:cosmological_models}

In this paper we study a number of cosmological models and compare their predictions given complementary cosmological data sets. While the choice of these models is highly biased, they each represent a larger class of cosmological models which attempt to solve or at least address some of the mentioned issues associated with the late-time acceleration of the Universe.

\subsection{GR-based cosmologies}
\label{sec:GRcosmo}
The first cases we study are based on the standard cosmological model, as it is obtained from the field equations of General Relativity with the Friedmann-Lemaître-Robertson-Walker (FLRW) metric ansatz, which implements the assumption that the Universe is homogeneous and isotropic,
\begin{equation}\label{eq:Metric_ansatz}
    \D s^2 = - \D t^2 + a(t)^2 \left[ \frac{\D r^2}{1 - k\, r^2} + r^2 \D \Omega^2 \right],
\end{equation}
where $k = 0, \pm1$ represents a flat, or positive/negative curvature universe.\footnote{In the case of $k \neq 0$, $r$ should be thought of as a dimensionless radial coordinate, $r/r_0$, rescaled by the radius of curvature.} Plugging this ansatz into Einstein's field equations yields two dynamical equations for the scale factor
\begin{gather}
    \frac{\dot{a}(t)^2 + k}{a(t)^2} = \frac{8\pi\, G_N\, \rho(t) + \Lambda}{3}\quad \text{and} \label{eq:Friedmann}\\
    \frac{\ddot{a}}{a} = \frac{4\pi G_N}{3} \left[\rho(t) + 3 p(t)\right] + \Lambda \xrightarrow{\text{Eq.~\eqref{eq:Friedmann}}} \dot{\rho} = -3H (\rho + p)\,.
\end{gather}
The latter of these is nothing but a continuity equation for an ideal fluid with energy density $\rho$ and pressure $p$, and $H = \frac{\dot{a}}{a}$ is the Hubble rate. Defining the critical energy density $\rho_c = \frac{3H_0^2}{8\pi G_N}$, we can put this into the familiar form
\begin{equation}\label{eq:Hubble}
    H(z)^2 = H_0^2 \left[ \Omega_r (1+z)^4 + \Omega_m  (1+z)^3 + \Omega_k  (1+z)^2 + \Omega_\Lambda \right],
\end{equation}
where we have introduced the cosmic redshift $z = a^{-1} - 1$, and introduced the density parameters today
\begin{equation}\label{eq:def_omegas}
    \Omega_{(m/r)} = \frac{\rho_{(m/r)}(t=t_0)}{\rho_c},\quad \Omega_k = - \frac{k}{H_0^2},\quad \Omega_\Lambda = \frac{\Lambda}{3H_0^2}\,.
\end{equation}
In these equations the label ($m$) refers to non-relativistic matter with an equation of state $p=0$, and ($r$) refers to relativistic degrees of freedom, i.e.~radiation with $p = 1/3 \rho$.

\subsubsection{Model predictions}

We now specify the models of interest.

\paragraph{Flat \LCDM~cosmology}
The simplest model we study is the concordance cosmology, i.e.~the FLRW metric with vanishing spatial curvature, $k = 0$, and a (positive) cosmological constant. This is described by the equation
\begin{equation}
    H(z)^2 = H_0^2 \left[ \Omega_r (1+z)^4 + \Omega_m  (1+z)^3 + \Omega_\Lambda \right] \,,
\end{equation}
which today ($z=0$ and $H(0) = H_0$) implies that $\Omega_\Lambda = 1 - \Omega_r - \Omega_m$ must hold.

\paragraph{\LCDM~with curvature}
A less minimal version of \LCDM\ is obtained by allowing for spatial curvature, cf.~Eq.\eqref{eq:Hubble}, which we will refer to as $k$\LCDM. It is described by the Friedmann equation~\eqref{eq:Hubble} together with the more general constraint
\begin{equation}
    \Omega_k = 1 - \Omega_\Lambda- \Omega_r - \Omega_m \,.
    \label{eq:oLCDM}
\end{equation}

\paragraph{Dynamical dark energy, $w$\LCDM}\label{sec:wLCDM}
Finally, we can be even more general by both dropping the requirement of spatial flatness and modifying the equation of state of the dark energy component, $p = w \rho$ which allows for an accelerated expansion as long as $w<-1/3$, thus representing a larger class of cosmological models of dynamical dark energy. One finds that
\begin{equation}
    H(z)^2 = H_0^2 \left[ \Omega_r (1+z)^4 + \Omega_m  (1+z)^3  + \Omega_k  (1+z)^2 + \Omega_\Lambda (1+z)^{3(w+1)} \right],
\end{equation}
and Eq.~\eqref{eq:oLCDM} must be satisfied, too.

\subsection{Bigravity cosmology}
\label{sec:bigra_cosmology}

Bigravity is a generalisation of the de Rham-Gabadaze-Tolley theory of ghost-free massive gravity, the theory of a massive spin-2 field, which requires the introduction of an auxiliary tensor field. In bigravity, this auxiliary tensor is dynamical itself, such that gravity is described by two tensor fields $g$ and $f$, which are coupled via a potential. The form of this potential is strongly constrained due to consistency requirements of the theory. In the parametrisation we will use, the bigravity action of our choice reads
\begin{align}
S_\text{bi}   =& \, \frac{M_g^2}{2} \int d^4 x \sqrt{-  g} \ R(g) +m^2 M_g^2 \int d^4 x  \sqrt{- g}  \sum_{n=0}^4 \beta_n e_n(\sqrt{g^{-1} f})\nonumber\\
&+\frac{M^2_{f}}{2} \int d^4 x \sqrt{-  f} \ \tilde{R}(f) +  \int d^4 x  \sqrt{-  g}  \,\, \mathcal{L}_{\text{matter}
},
\label{eq:bigra_action}
\end{align}
where the bigravity potential is built from elementary symmetric polynomials~\cite{Hassan:2011vm}, and the corresponding equations of motion are
\begin{subequations}
\label{eq:Einstein}
\begin{align}
&R_{\mu\nu} -\frac{1}{2} g_{\mu\nu} R + B_{\mu\nu}(g,f) = \frac{8 \pi}{M_g^2} T_{\mu\nu}\,,\\
&\tilde{R}_{\mu\nu} -\frac{1}{2} f_{\mu\nu} \tilde{R} + \tilde{B}_{\mu\nu}(g,f) = 0\,,
\end{align}
\end{subequations}
where $R_{\mu\nu}$ ($\tilde{R}_{\mu\nu}$) is the Ricci tensor constructed from $g$ ($f$), while the $B_{\mu\nu}$ and $\tilde{B}_{\mu\nu}$ are derived from the potential $V(g,f)$ (see \cite{vonStrauss:2011mq} for explicit expressions). Here, $M_g$ is the Planck mass corresponding to the physical metric $g$, while the auxiliary metric  comes with the mass scale~$M_f$.

Making $f$ a dynamical field has a number of advantages, foremost it removes the arbitrariness of the reference metric, which instead obeys a dynamical field equation.\footnote{While we focus on bigravity, note that there are viable cosmological solutions in  massive gravity for an appropriate, non-flat reference metric~\cite{DeFelice:2012mx,Comelli:2012db}.} In the setup of bigravity which we investigate here, where matter couples exclusively to the physical metric $g$, while the other tensor $f$ is regarded as an additional degree of freedom rather than a geometrical object. Under these assumptions and a bi-FLRW ansatz~\cite{PhysRevLett.119.111101}, the equations governing the dynamics of the universe read
\begin{subequations}\label{eq:BiFriedmann}
\begin{align}
	\frac{3}{a^2}\left(H^2+k \right) - m^2  \left[ \beta_0 + 3\beta_1 y + 3\beta_2 y^2 + \beta_3 y^3 \right] &= 8 \pi\,G_N \rho, \label{eq:Einstein_g}\\
	\frac{3}{b^2}\left(J^2/\tilde{c}^2+k \right) - m^2 \frac{1}{\alpha^2} \left[ \beta_1 y^{-3} + 3\beta_2 y^{-2} + 3\beta_3 y^{-1} + \beta_4 \right]&=0.\label{eq:Einstein_f}
\end{align}
\end{subequations}
with $\alpha \equiv \frac{M_f}{M_g}$.
Here $a$ is the scale factor of the physical metric and $b$ that of the auxiliary metric, while $\tilde{c}$ is the lapse of the auxiliary metric. We have also defined $J\equiv \frac{\dot{b}}{b}$ and  $y \equiv b/a$. The parameters $\beta_i$ are constants of a~priori unknown magnitude, while $m$ is a mass scale related to the physical graviton mass (see below). 
It is evident from Eqs.~\eqref{eq:BiFriedmann} that the parameters $\beta_i$ and the mass scale $m$ are not independent parameters; however, the latter is conventionally factored out by introducing a new set of parameters,
\begin{align}
\qquad \qquad B_i \equiv \beta_i\, \frac{m^2}{H_0^2} \qquad \quad i=0,1,2,3,4
\end{align}
{for reasons that will become apparent momentarily. Moreover, it can be shown from the action that a rescaling of the hidden sector Planck mass, $M_f$, can always be compensated by an appropriate rescaling of the $B_i$. Thus, the ratio $\alpha$ is also not an independent parameter~\cite{Luben:2020xll}. For our statistical analysis, we will set $\alpha = 1$ and choose an appropriate range for the sampling of the $B_i$; for details on our choice of priors on the bigravity parameters, we refer to Sec.~\ref{sec:Bigra_results}. We stress that a rescaling invariance remains for all cosmological solutions which we will present, under which $B_i \mapsto \alpha^{-i} B_i$~\cite{Luben:2020xll}.

As a related note, we briefly discuss the role of perturbations on the cosmological background. A number of analyses on the topic exist for bigravity~\cite{Berg:2012kn,Konnig:2014dna,Solomon:2014dua,Konnig:2014xva,Lagos:2014lca,Cusin:2014psa,Enander:2015vja,Cusin:2015pya,Kobayashi:2015yda,Lagos:2016gep}; however, it is generally found that the models suffer from gradient or ghost instabilities at the linear level of perturbations~\cite{Amendola:2015tua,Konnig:2015lfa}, which can either be avoided by fine tuning the initial conditions~\cite{Johnson:2015tfa}, or by considering more exotic matter couplings~\cite{Comelli:2012db}. It has also been argued by some authors that these instabilities could actually serve as seeds for structure formation by virtue of the Vainshtein mechanism that is expected to set in when these instabilities lead to large overdensities~\cite{Mortsell:2015exa}.

Finally, there is also the viewpoint that it could be sufficient to shift the instabilities of the cosmological solution to the very early Universe. In the conventional parametrisation with $\alpha$ as a free parameter, taking $\alpha = \order{10^{-17}}$ and choosing the $B_i$ to be of the same order can push the age of the onset of instabilities up to the time of Big Bang nucleosynthesis (BBN)~\cite{Akrami:2015qga}. This parameter space is therefore a region of particular interest, and is also the limit which recovers GR from bigravity. However, this feature is obscured in our minimal choice of parametrisation: as we pick $\alpha = 1$, the $B_i$ need to be tuned finely over several different orders of magnitude in order to reproduce the same type of background cosmology.}

Notice that \eqref{eq:Einstein_g} is identical to the standard Friedmann equation augmented by a dynamical CC,
\begin{equation}
\label{eq:bigra_CC}
\Lambda(z) = H_0^2 \left[ B_0 + 3 B_1 y(z) + 3B_2 y^2(z) + B_3 y^3(z) \right],
\end{equation}
which is now a function of redshift $z$ and thus time. 

\paragraph{Determination of the ratio of scale factors.}
The specific form of coupling the two tensor fields implies that several branches of solutions of Eq.~\eqref{eq:BiFriedmann} exist. We choose the dynamic/finite branch~\cite{vonStrauss:2011mq}, for which $J^2/\tilde{c}^2=H^2$. This allows us to rewrite the two Friedmann equations into one \textit{master equation} for the ratio of the two scale factors,
\begin{equation}
\label{eq:bigra_master_eq}    
y^4 + a_3 y^3 + a_2 y^2 + a_1(z) y + a_0 =0
\end{equation}
with
\begin{align*}
a_3&=\frac{B_4 -3B_2\, \alpha^2}{-B_3 \, \alpha^2} \quad & \quad  a_2&=\frac{3B_3 -3B_1 \, \alpha^2}{-B_3 \, \alpha^2} \nonumber\\
a_1(z)&=\frac{(3B_2 -B_0\, \alpha^2)- 3\,( \Omega_m \,(1+z)^3 + \Omega_r \,(1+z)^4 + \Omega_\Lambda)  \, \alpha^2}{-B_3 \, \alpha^2} \quad & \quad  a_0&=- \frac{B_1}{B_3} \frac{1}{\alpha^2} \nonumber\\
\end{align*}
where we have replaced the energy density $\rho$ with the matter density parameter $\Omega_m$ and radiation component $\Omega_r$ as well as a constant dark energy component $\Omega_\Lambda$.\footnote{Note that $\Omega_\Lambda$ and $B_0$ are degenerate; therefore, in our numerical analysis we have chosen $B_0$ such that Eq.~\eqref{eq:bigra_CC} yields no additional constant dark energy component at $z=0$. Hence, $\Omega_\Lambda$ has the expected physical interpretation.} Note that $a_1$ brings in a redshift-dependence. In particular, for $z \to \infty$, it scales as $a_1 \to \pm \infty$.

In order for our chosen background solution to be viable, we enforce the model to lie on the \emph{finite branch} of solutions, i.e.~that $y$ vanishes at $z\to\infty$ and evolves to a finite value in the distant future, $z=-1$. While it is possible to have a solution that yields $y\to\infty$ in the early universe, such \emph{infinite branch} solutions are unphysical~\cite{vonStrauss:2011mq,DeFelice:2014nja}.
This can be seen by taking a derivative w.r.t.~$\log a$ of Eq.~\eqref{eq:bigra_master_eq} and recasting it into~\cite{Luben:2020xll}
\begin{equation}
  \frac{\D y}{\D \log a} = y \frac{(1+w_m) \rho_m / M_g^2}{m_\text{eff}^2 - 2H^2},
\end{equation}
where
\begin{equation}\label{eq:yprime}
  m_\text{eff}^2 \equiv \left(1 +   \frac{1}{\alpha^2\,y^2} \right) H_0^2 (y\,B_1 + 2 y^2 B_2 + y^3 B_3) 
\end{equation}
is the time-dependent, effective spin-2 mass. This quantity musst satisfy the Higuchi bound $m_\text{eff}^2 \ge 2H^2$ at all times, as otherwise the ghost DOF re-appears~\cite{Higuchi:1986py}. Thus, in the infinite branch, where for $z\to\infty$, $y \to \infty$, we have $y' < 0$ in order to end the evolution at a finite value for $y$ today. Thus, we must either violate the Higuchi bound, which renders the theory inconsistent, or we must introduce exotic forms of matter with $\rho_m < 0$, which dominate the universe.

Conversely, Eq.~\eqref{eq:yprime} highlights that, when considering the finite branch and a regime where $\rho_m >0$, $m_\text{eff}^2 \ge 2H^2$, $y'$ cannot change signs, i.e.~the asymptotic value $y_* = y(z\to \infty)$ fixes the sign of $y$ for all redshifts. Since we must have $H(z)^2 \ge 0$ for all $z$, this equation further implies that $m_\text{eff}^2 \ge 0$, and thus~\cite{Luben:2020xll}
\begin{equation}
   B_1 > 0
\end{equation}
in the finite branch.	

We now continue our approach to solving Eq.~\eqref{eq:bigra_master_eq}. The solutions to this fourth order equation read
\begin{equation}
\label{eq:bigra_sols}
    y_{1,2} = - \frac{a_3}{4} + \frac{R}{2} \pm \frac{D}{2} \qquad
    y_{3,4} = - \frac{a_3}{4} - \frac{R}{2} \pm \frac{E}{2} \nonumber
\end{equation}
with
\begin{subequations}
\begin{align}
    R &\equiv \sqrt{\frac{1}{4}a_3^2 -a_2 +x_1(z)}\,,\\
    D &\equiv \sqrt{\frac{3}{4} a_3 - R^2-2\,a_2+\frac{1}{4}(4\, a_3\,a_2-8a_1-a_3^3)R^{-1}}\,,\\
    E &\equiv \sqrt{\frac{3}{4} a_3 - R^2-2\,a_2 - \frac{1}{4}(4\, a_3\,a_2-8a_1-a_3^3)R^{-1}}\,,
\end{align}
\end{subequations}
and where $x_1$ is a real root of
\begin{equation}\label{eq:cubic}
    x(z)^3 - a_2 x(z)^2 + [a_1(z) a_3 -4 a_0] x(z) + [4 a_2 a_0 - a_1^2(z) - a_3^2 a_0] = 0.
\end{equation} 
Recall that we require that $y \to 0$ for $z\to\infty$, which is enforced by the infinite energy density limit in the early Universe~\cite{vonStrauss:2011mq}. Let us therefore have a look at the asymptotic behaviour of Eq.~\eqref{eq:cubic}, where by assumption $|a_1(z)| \gg |a_{0,2,3}|$, and keeping all orders in $x(z)$:
\begin{equation}
    x(z)^3 - a_2 x(z)^2 + a_1(z) a_3\, x(z) - a_1^2(z) = 0\,,
\end{equation}
which has only one real solution, which scales as 
\begin{equation}
    x(z \to\infty) \equiv x_\infty = \frac{a_2}{3} + a_1(z)^\frac{2}{3} - \frac{a_3}{3}\, a_1(z)^\frac{1}{3} 
\end{equation}
if $B_3 > 0$ and
\begin{equation}
    x(z \to\infty) \equiv x_\infty = \frac{a_2}{3} + |a_1(z)|^\frac{2}{3} + \frac{a_3}{3}\, |a_1(z)|^\frac{1}{3} 
\end{equation}
in the case $B_3 < 0$. Thus, we always have that asymptotically $R_\infty = \sqrt{x_\infty}$ and in the limit of large redshift, 
\begin{equation}
D_\infty = \sqrt{- x_\infty - \frac{2\, a_1(z)}{\sqrt{x_\infty}}},\  E_\infty = \sqrt{- x_\infty + \frac{2\, a_1(z)}{\sqrt{x_\infty}}}\,.
\end{equation}
It is clear from this equation that only one solution can be asymptotically real, either involving $D$ ($a_1<0$) or $E$ ($a_1>0$), where the sign of $a_1$ is fixed by $B_3$. For a parametric scan, we pick a sample of the parameters $B_i$ and use the asymptotically real branch. In summary, we have identified the unique solution branch only by demanding that GR be restored at sufficiently early times.

In order to comply with our assumption that $y\to 0$ for $z \to \infty$, we must then also demand that $a_3 = 0$. This is consistent with the requirement $d \rho / dy < 0$, a condition sufficient to avoid the Higuchi ghost instability~\cite{Yamashita:2014cra,DeFelice:2014nja,Konnig:2015lfa}.

Finally, we determine the physical graviton mass by taking the future limit, where $z=-1$ and $y$ goes to a constant value $y_*$. The graviton mass is then~\cite{Max:2017kdc}
\begin{equation}
\label{eq:bigra_graviton_mass_mg}
   \qquad  m_g^2 = y_* H_0^2\, G_* \qquad \text{with}\quad G_* \equiv (B_1+2y_* B_2 + y_*^2 B_3)=\text{const.}
\end{equation}
The future fixed point of the scale factor $y_*$ is determined from the master equation~\eqref{eq:bigra_master_eq} for $\rho \to 0$.

By having chosen the finite branch, we obtain a stable cosmology at the background level. However, linear perturbations on this background solution are known to develop a scalar instability at early times~\cite{Comelli:2012db,Konnig:2014xva,Konnig:2015lfa,Lagos:2016gep}. While this instability occurs at the perturbative level, it is currently unclear whether this issue is resolved due to strong coupling effects; the on-set of the instability may also be pushed to unobservable early times, which can be achieved by a large hierarchy of the Planck masses~\cite{Akrami:2015qga}.

We conclude this section with a sketch of the cosmic history in bigravity: In the early Universe, dominated by matter and radiation, the modifications of bigravity are irrelevant, as we have chosen a finite branch solution with $y \to 0$ for $z \to \infty$; this is identical to a \LCDM~cosmology without any CC contribution (which is irrelevant at large $z$). At a certain redshift, $y$ will develop dynamics and modify the expansion history of the Universe, effectively through a dynamical CC, cf.~Eq.~\eqref{eq:BiFriedmann}. Finally, the scale factors reach a constant ratio, $y_*$, which is the value assumed in the far future, $z=-1$. If we live in a bigravity universe, where we are sufficiently far away from this equilibrium point, we may hope to identify the characteristic features of the dynamical CC in this model, or constrain it otherwise. 

Notice the similarity between this behaviour and the behaviour found in spherically symmetric solutions in bigravity~\cite{Platscher:2016adw}. In this metric space, the potential looks Newtonian far away from the source; however, at a certain distance from the source $r_c = m_g^{-1}$, the solution begins to deviate from GR and develops a Yukawa-type potential. Finally, and even closer to the source at a distance $r_V$, the longitudinal polarisation modes of the massive spin-2 field will become strongly coupled and non-linearities conspire to restore the GR predictions by rendering any longitudinal polarisation state non-dynamical~\cite{Vainshtein:1972sx}. This so-called Vainshtein screening is indeed also incorporated in the cosmological solution we employ, which was obtained without any assumptions about linearity.
See also Ref.~\cite{Luben:2019yyx} for a recent study of this effect in cosmology.

\subsection{Conformal gravity cosmology}

\label{sec:CG}
CG is a generalization of GR, that demands conformal symmetry in addition to general covariance.  The CG action,
 \begin{equation}
    S_\text{CG} = - \alpha_g \int \D^4x \sqrt{-g}\, C_{\lambda\mu\nu\kappa} C^{\lambda\mu\nu\kappa}\,,
    \label{eq:CG_action}
\end{equation}
is constructed from the Weyl tensor $C^\lambda_{~\mu\nu\kappa}$, which is the complete traceless part of the Riemann tensor\footnote{The Weyl tensor is defined by $ C_{\lambda\mu\nu\kappa}=R_{\lambda\mu\nu\kappa}-(g_{\lambda[\nu}R_{\kappa]\mu}-g_{\mu[\nu}R_{\kappa]\lambda}) + \frac{1}{3} g_{\lambda[\nu} g_{\kappa]\mu} R$.} and is conformally invariant. By construction, the coupling constant $\alpha_g$ is dimensionless and the action is invariant under conformal transformations where the metric is locally rescaled by\footnote{We adopt the convention with (--,+,+,+) metric signature and the Riemann tensor defined by $R^\mu_{\nu\alpha\beta} = \Gamma^\mu_{\nu \beta, \alpha} + \Gamma^\mu_{\sigma \alpha} \Gamma^\sigma_{\nu\beta} - (\alpha \leftrightarrow \beta)$.}
\begin{equation} \label{eq:conftransf}
    g_{\mu\nu}(x) \to \Omega(x)^2 g_{\mu\nu}(x)\,.
\end{equation}
Due to quadratic dependence on curvature invariants in Eq.~\eqref{eq:CG_action}, the actions depends on up to fourth-order derivatives of the metric, a fact which can be seen as a virtue and as a disadvantage. On the one hand, the inclusion of these higher-order terms renders these theories renormalizable by naive power-counting arguments~\cite{PhysRevD.16.953}. On the other hand, these terms give rise to new degrees of freedom containing a spin-2 ghost state \cite{Stelle:1977ry,Riegert:1984hf}. Such a degree of freedom is in general considered unphysical since it suffers from the so-called Ostrogradski instability at the classical level~\cite{Ostrogradsky:1850fid}, and consequently unitarity is violated in the quantum theory~\cite{Woodard:2015zca}. However, proposals exist to deal with the ghost state  (see e.g. Refs.~\cite{Bender:2007wu,Mannheim:2018ljq,Salvio:2019ewf,Shapiro:2014fsa,Donoghue:2019fcb,Salvio:2015gsi, Raidal:2016wop, Hawking:2001yt}). In this work, we intend to study a particular cosmological model following the ideas of Refs.~\cite{Mannheim:1989jh, Mannheim:2005bfa} and references therein. In addition, this model offers a possible solution to the missing mass problem of galaxies which we discuss in Sec.~\ref{sec:SWinCG}. In the past, attempts have been made to explain galactic rotation curve data without the addition of a dark matter halo in this model~\cite{Mannheim:2012qw}, and furthermore, Refs.~\cite{OBrien:2017gam, OBrien:2017xaw} also address observed galaxy cluster motion with no dark matter. Furthermore, if CG is to account for all dark matter in the Universe, it is so far unclear if it can pass gravitational lensing tests \cite{Cutajar:2014gfa,Kasikci:2018mtg,Sultana:2010zz, PhysRevD.58.024011,Cattani:2013dla,Lim:2016lqv,Campigotto:2017ytw}, and inconsistencies with gravitational wave observation of binaries have been found~\cite{Caprini:2018oqe}. Also, tensions between predictions of primordial nucleosynthesis in a CG cosmology and observation of light element abundances have been found in Refs.~\cite{Knox:1993fj,Elizondo:1994vh}.
In the remainder of this section we discard all these concerns for now and review the derivation of the modified Friedmann equations in CG following closely Ref.~\cite{Mannheim:2005bfa}.
 The field equations obtained from Eq.~\eqref{eq:CG_action}, also known as Bach equations~\cite{Bach1921}, read
\begin{equation}\label{eq:Bach}
    4 \alpha_g W_{\mu\nu} \equiv 4 \alpha_g \qty(2 \nabla^\alpha \nabla^\beta C_{\alpha\mu\nu\beta} + C_{\alpha\mu\nu\beta} R^{\alpha \beta})  = T_{\mu\nu}\,,
\end{equation}
where the Bach tensor $W_{\mu\nu}$ can be understood as the generalization of the Einstein tensor and the energy-momentum tensor $T_{\mu\nu}$ can be derived from a conformally invariant matter action, e.g.~containing a complex scalar $\phi$ and a fermion $\psi$,
\begin{equation} \label{eq:matter_action}
    S_\text{M} = - \int \D^4x \sqrt{-g}\, \left[ \left(\frac{1}{2} (\nabla^\mu\phi)^\dag \nabla_\mu \phi + \frac{R}{6} \phi^\dag \phi \right) + \lambda (\phi^\dag \phi)^2 + i \overline{\psi} \slashed{D} \psi + y\phi \overline{\psi}\psi \right] \,.
\end{equation} 
Due to conformal invariance, the non-minimal coupling term, $R\, \phi^\dag\phi$, is required and it introduces a piece proportional to the Einstein tensor $G_{\mu\nu}$ in the energy-momentum-tensor~(EMT)
\begin{equation} \label{eq:EMT}
    T_{\mu\nu} = T_{\mu\nu}^\text{GR} + \frac{1}{6} \phi^\dag \phi\, G_{\mu\nu} \,,
\end{equation}
where $T_{\mu\nu}^\text{GR}$ is the usual matter EMT.

Amusingly, the FLRW ansatz for the metric [cf. Eq.~\eqref{eq:Metric_ansatz}] is conformally indistinguishable from a flat solution which satisfies the vacuum equation $W_{\mu\nu}=0$, and thus the Bach equation reduces to Einstein's field equations with a flipped sign $\expval{\phi^\dag \phi} G_{\mu\nu} = -6 \, T_{\mu\nu}^\text{GR}$ and an effective gravitational coupling constant set by the vacuum expectation value (VEV) of the field~$\phi$. Whether or not we wish to view the field~$\phi$ as the Higgs field that breaks both electroweak and conformal symmetry, once it takes a constant field value, it will break the conformal symmetry and set the scale of gravitational interactions. Plugging the FLRW metric ansatz \eqref{eq:Metric_ansatz} into the field Eq.~\eqref{eq:Bach} and assuming that $T_{\mu\nu}^\text{GR}$ constitutes a perfect fluid leads to the modified Friedmann equation of CG,
\begin{equation}\label{eq:Friedmann_2}
H(z)^2 = - \epsilon \, H_0^2 \left[ \Omega_r (1+z)^4 + \Omega_m (1+z)^3 + \Omega_\Lambda \right] + H_0^2 \Omega_k (1+z)^2,
\end{equation}
where the densities $\Omega_{m,r,k}$ are defined as in Eqs.~\eqref{eq:def_omegas} but the dark energy density is set by the VEV of the scalar $\Omega_\Lambda = \lambda \expval{\phi^4}/H_0^2$. In Eq.~\eqref{eq:Friedmann_2} we have introduced the dimensionless quantity $\epsilon \equiv \frac{3}{4\pi\,G_N\, \langle\phi^2 \rangle}$. This allows us to define \emph{modified} energy densities $\overline{\Omega}_i = - \epsilon\, \Omega_i$\footnote{If we wish to achieve $\overline{\Omega}_\Lambda > 0$, we must accept that $\lambda < 0$ in the action, since $\rho_\Lambda = \lambda \langle\phi^4\rangle$.}
and bring the Friedmann equation into the familiar form
\begin{equation}\label{eq:Friedmann_3}
\frac{H(z)^2}{H_0^2} = \left[ \overline \Omega_r (1+z)^4 + \overline \Omega_m (1+z)^3 + \overline \Omega_\Lambda \right] +  \Omega_k (1+z)^2 \,,
\end{equation}
where matter and radiation contribute negatively to the Friedmann equation, as $\overline{\Omega}_{(m)r}<0$, and the cosmological constant is assumed to contribute positively as demanded by observations.\footnote{For $\Omega_k > 0$, which is demanded by galactic rotation curves (see Sec.~\ref{sec:SWinCG}), the Universe is in an accelerated phase at all times.} It should be stressed that the physical densities of matter $\rho_m$ and $\rho_r$ are still positive, but their gravitational interactions are repulsive on cosmological scales.

Due to the negative energy density parameters entering Eq.~\eqref{eq:Friedmann_3}, there is a \emph{maximal} redshift $z_\text{max}$, which is reached once the squared Hubble rate has a root, $H(z_\text{max}^2) = 0$. For a flat universe with only matter (radiation) and a cosmological constant, the maximal redshift is 
\begin{equation}
z_\text{max} \approx \left(\frac{\overline{\Omega}_\Lambda}{- \overline{\Omega}_{m(r)}}\right)^{\frac{1}{3}\, \left(\frac{1}{4} \right)}\,,
\end{equation}
which translates to a minimal scale factor $a_\text{min}=1/(1+z_\text{max})$.

For the present analysis we adapt the following assumptions, which allow us to put conservative bounds on CG. Processes in the early Universe such as recombination and nucleosynthesis are well established via the CMB and the abundance of nuclei in the Universe, respectively. Therefore, we demand that $z_\text{max} > 10^{15}$, for BBN to be safely inside the expanding phase when these processes take place. Thus, $\overline{\Omega}_{m(r)}$ must be tiny if we assume $\overline{\Omega}_\Lambda \sim 1$, which can only be achieved if $\epsilon < 10^{-15}$. 

We also take into account the vacuum energy due to the scale of the electroweak phase transition. To this end, we must bring into agreement the observed vacuum energy density $\rho_\text{vac}^\text{obs} \sim (10^{-2}\eV)^4$ with the expected contribution $\rho_\text{vac}^\text{EW} \sim (100\GeV)^4$. For that contribution we have $\overline{\Omega}_\Lambda = \epsilon \rho^\text{EW}_\text{vac}/\rho_c \sim 10^{54} \epsilon $, so the appropriate order of magnitude suppression demands that $\epsilon \sim 10^{-54}$. Furthermore, we have to consider the contribution coming from the VEV of the scalar $\phi$ to the cosmological constant $\rho_\Lambda = \lambda \expval{\phi^4}$. The VEV $\expval{\phi}$ is set by $\epsilon \sim 10^{-54}$ to $\expval{\phi} \sim 10^{26} M_\text{Pl}$. This huge vacuum expectation value requires to fine-tune the scalar self-coupling to $\lambda \sim 10^{-176}$ in order to maintain $\overline{\Omega}_\Lambda \sim 1$. This vast amount of fine-tuning indicates that the cosmological constant problem persists in CG. However, we find that $\epsilon$ is constrained even stronger due to vacuum energy contributions to $\overline\Omega_\Lambda$ so that the BBN constraint is easily satisfied.

From the above considerations we conclude that $\epsilon \ll 1$, and that only the vacuum energy and curvature (the latter of which is not multiplied by $\epsilon$) contribute to the cosmology of CG in the range of redshifts we are interested in,
\begin{equation}
\label{eq:CGlowz}
H(z)^2 = H_0^2 \left[ \overline{\Omega}_\Lambda + \Omega_k (1+z)^2 \right] \,.
\end{equation}

\subsubsection{Galactic Rotation Curves without Dark Matter}
\label{sec:SWinCG}
As mentioned above, the modified gravitational potential of CG has been used to address the missing mass problem of galaxies in the past~\cite{Mannheim:1988dj,Mannheim:1992tr,Mannheim:1992vj,Kazanas:1988qa}. Here, we briefly review the potential generated by a spherical symmetric source and novel effects that arise only in CG following closely Ref.~\cite{Mannheim:2012qw}. The potential generated outside the source of radius $R$ reads
\begin{equation}
\label{eq:localpot}
	V^*(r>R)= - \frac{\beta^*}{r} + \frac{\gamma^* r}{2} \,.
\end{equation}
For small radius $r$ the Newtonian limit is recovered if $\beta^* = G_N M$, and the linear term in $r$ marks a departure from the known behaviour on larger scales. Due to the fourth-order derivatives inherent to CG, Newton's shell theorem is no longer valid.
The global contribution can be divided into two components: the homogeneous cosmological background and the inhomogeneities on this background. First, we consider the homogeneous and isotropic background described by the FLRW metric~\eqref{eq:Metric_ansatz}. To compute the gravitational potential due to the ambient FLRW background which an observer in the Schwarzschild rest frame experiences, one can use general coordinate invariance. By a suitable coordinate transformation for the time and radial coordinates, Eq.~\eqref{eq:Metric_ansatz} is transformed into a Schwarzschild-type metric
\begin{equation}\label{eq:Schwarzschild}
\D s^2 = \Omega^2(\tau,\rho) \qty[-(1+\gamma_0 r) \D t^2 + \frac{\D r^2}{(1+\gamma_0 r)} + r^2 \D \Omega^2] \,.
\end{equation}
This reveals that the FLRW background generates the linear term in the potential for non-zero curvature, and this term is related to the spatial curvature $k$ through the relation\footnote{For a well-behaved potential we demand that $k<0$ which corresponds to an open universe with $\Omega_k > 0$.}
\begin{equation}
\gamma_0 = 2 \sqrt{-k}. 
\label{eq:gammakrelation}
\end{equation}
 This allows us to test the parameter $\gamma_0$ on two distinct scales. On the one hand, it appears as a global term in the potential below, where it can be determined by local data such as galactic rotation curves. On the other hand, $k$ will enter the Friedmann Eq.~\eqref{eq:CGlowz} through the curvature density $\Omega_k$ and its value can be constrained by a cosmological fit. We discuss this particular feature in Sec.~\ref{sec:CG_results}.
 
 The second, global contribution to the potential is due to
inhomogeneities on the FLRW background, which introduce a quadratic term in the potential (see Ref.~\cite{Mannheim:2012qw} for details). To sum up, we obtain for the full potential outside the source
\begin{equation}
\label{eq:fullpot}
V(r > R) = V^*(r>R) + \frac{\gamma_0 r}{2} - \kappa \, r^2,
\end{equation}
from which galactic rotation curves are predicted in Refs.~\cite{Mannheim:2012qw,Mannheim:2019aap, Mannheim:2010ti, Mannheim:2010xw, OBrien:2011vks, OBrien:2017bwr}. 

\newpage


\section{Cosmological fits}\label{sec:results}
In this section we present the results of the numerical data analyses in the various cosmological models. For each model, the full list of sampling parameters consists of the model-specific parameters introduced in Sec.~\ref{sec:cosmological_models}, plus a number of universal nuisance parameters, which we describe in Appendix~\ref{app:data}.
First off, we study the flat concordance cosmology, \LCDM. We then turn to two parametrised extensions of \LCDM, adding curvature ($k$\LCDM) and finally a free DE equation of state ($w$\LCDM) in Secs.~\ref{sec:oLCDM} and~\ref{sec:wLCDM_results}. 
Here, we will validate our findings in the context of existing surveys, and we will present for the first time results of a combined analysis using SNe, BAO, CMB data in conjunction with the advertised quasar standard candles.  
In order to compare the validity of the models, we give the so-called Bayes information criterion (BIC), which is defined in the Appendix.

Subsequently, we turn to non-parametric extensions of \LCDM, the first of which is bigravity --~with and without curvature. 
Finally, we discuss the phenomenological implications within CG, where we will conclude that this framework is not apt to explain the current data, and can essentially be ruled out in its basic formulation. Nevertheless, we also show that an intriguing feature of the model is that certain parameters appear both in the cosmological solution as well as local geometries, and therefore both galactic and cosmological data can constrain the same set of parameters. However, it turns out that the two data sets yield incompatible results for the model parameters. We hope that these results can point towards a phenomenologically viable theory in the future.
We summarize our results in Table \ref{tab:results}.

\subsection{Flat \LCDM\ model}\label{sec:LCDM}

\begin{figure}[ht]
    \centering
    \includegraphics[width=\textwidth]{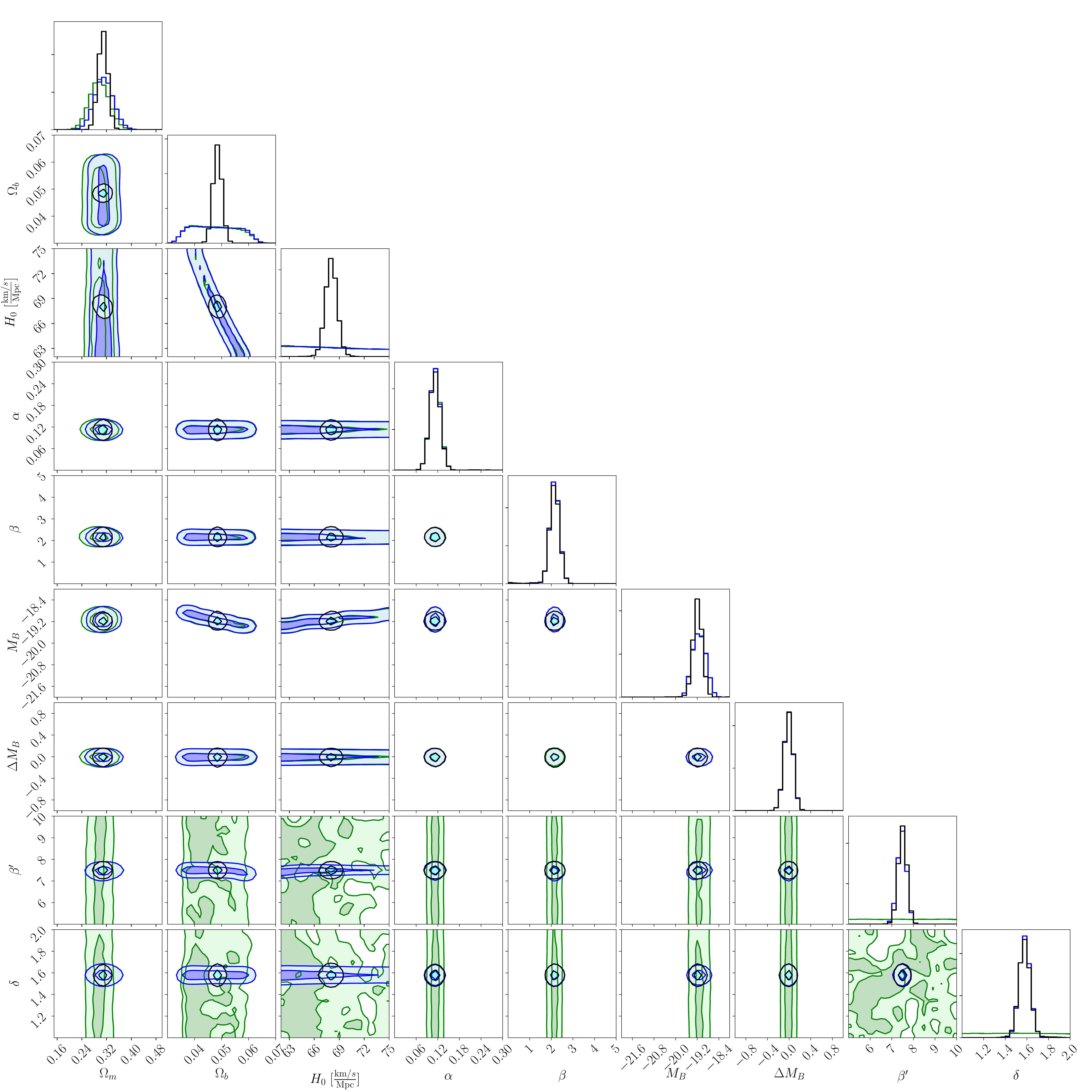}
    \caption{Marginalised 1D and 2D posterior distributions for~\LCDM~for all cosmological and nuisance parameters for SNe (green), SN+quasars (blue) and combined SN+quasars+BAO+CMB (black) data sets, also including the Gaussian prior on $\Omega_b \,h^2$. The SN data alone does not constrain the quasar nuisance parameters $\beta'$ and $\delta$, while the inclusion of quasars has only little effect on the SN nuisance parameters. Notice also that there is no significant correlation between the nuisance and cosmological parameters. See the Appendix for the definitions of all nuisance parameters.}
    \label{fig:LCDM_full_posterior}
\end{figure}

In a first step, we apply the techniques introduced in Appendix~\ref{app:data} to a standard \LCDM~model to cross validate our findings with the literature. Setting up a total of 512 uniformly sampled parameter points and evolving the Markov Chain Monte Carlo (MCMC) sampler for 1000 iterations, we find for different data sets the posterior distributions shown in the rightmost panels of Fig.~\ref{fig:LCDM_full_posterior}. Note that we have chosen flat priors whose allowed ranges encode some physical expectation, i.e.~$60\kmsmpc<H_0<80\kmsmpc$ and $0<\Omega_m<1$, and a Gaussian prior for $\Omega_b \,h^2$ (see App.~\ref{app:data}), which implements independent information from nucleosynthesis (which we thereby implicitly assume to proceed in a standard manner).\footnote{The radiation density is negligible at low redshifts, hence $\Omega_\Lambda +\Omega_m = 1$. We are then left with $\Omega_m$, $\Omega_b$ and $H_0$ as the cosmological parameters.} From Fig.~\ref{fig:LCDM_full_posterior} we can draw a number of important conclusions: First, there is no significant correlation between model and nuisance parameters. Second, SN nuisance parameters are only affected by SN data and do not respond significantly to the inclusion of quasar data and vice versa. Thus, the calibration can --~in principle~-- be done independently, and we see that the combined data sets (which also include BAO and CMB data) yield confidence intervals that are compatible with the individual analyses. Third, the $M_B$--$H_0$ panel of Fig.~\ref{fig:LCDM_full_posterior} shows that SN data (and also quasar data) alone cannot constrain $H_0$ as their absolute magnitudes are degenerate with $H_0$ --~even if only weakly. In order to calibrate the SN data, we need to break this degeneracy, e.g.~by measurements involving standard rulers, or $H(z)$ measurements as given by the BAO data, cf.~App.~\ref{app:data}. And finally, our findings are in agreement with those in the literature for the JLA SNe sample, cf.~Ref.~\cite{JLA}, and the quasar sample~\cite{Risaliti:2015zla,Melia:2019nev}.

\begin{figure}[t]
    \centering
    \hspace{-12mm}
    \begin{subfigure}[r]{.45\linewidth}
    \includegraphics[width = 1.1\textwidth]{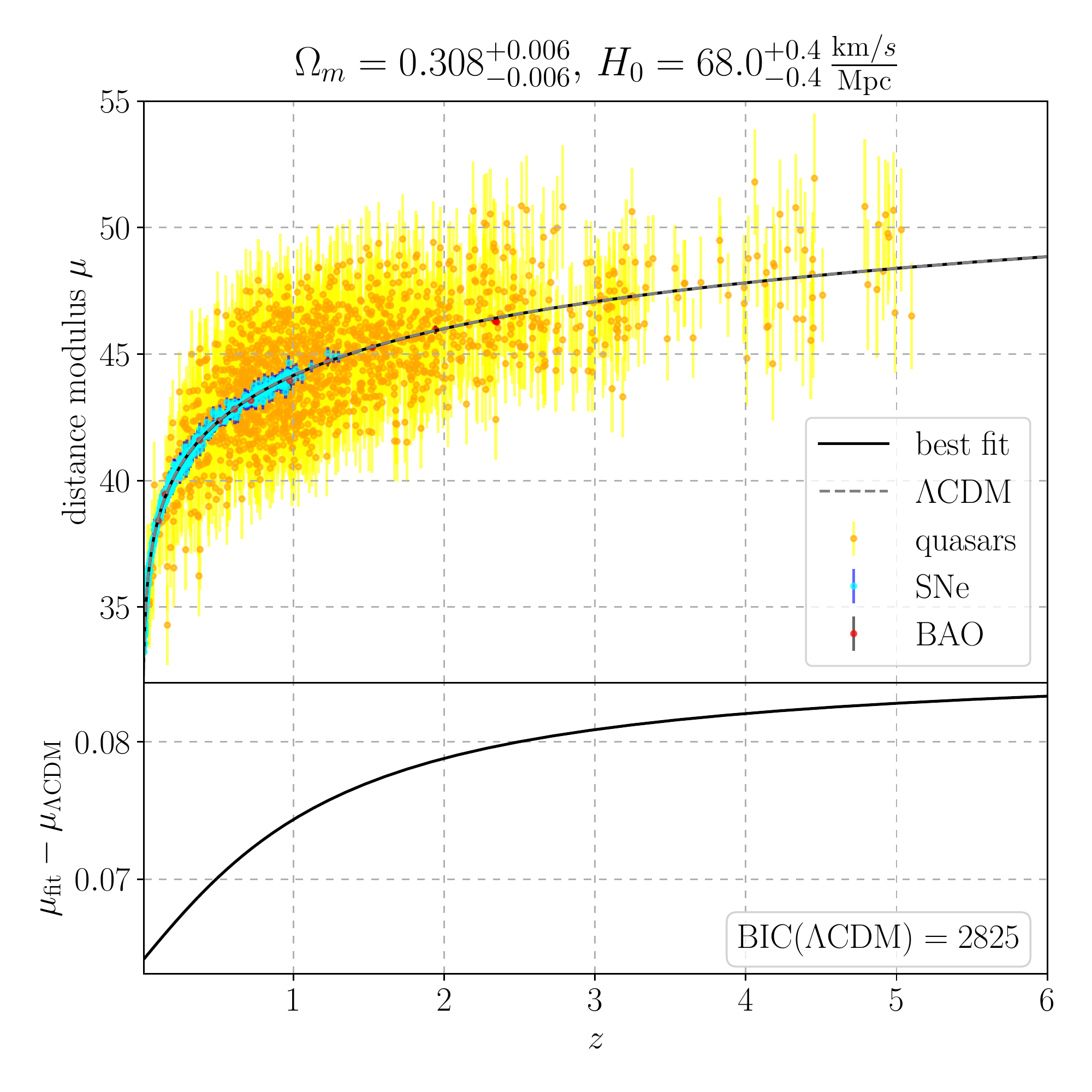}
    \subcaption{Hubble diagram in \LCDM}
    \end{subfigure}
    \hspace{8mm}
    \begin{subfigure}[l]{.45\linewidth}
    \includegraphics[width = 1.1\textwidth]{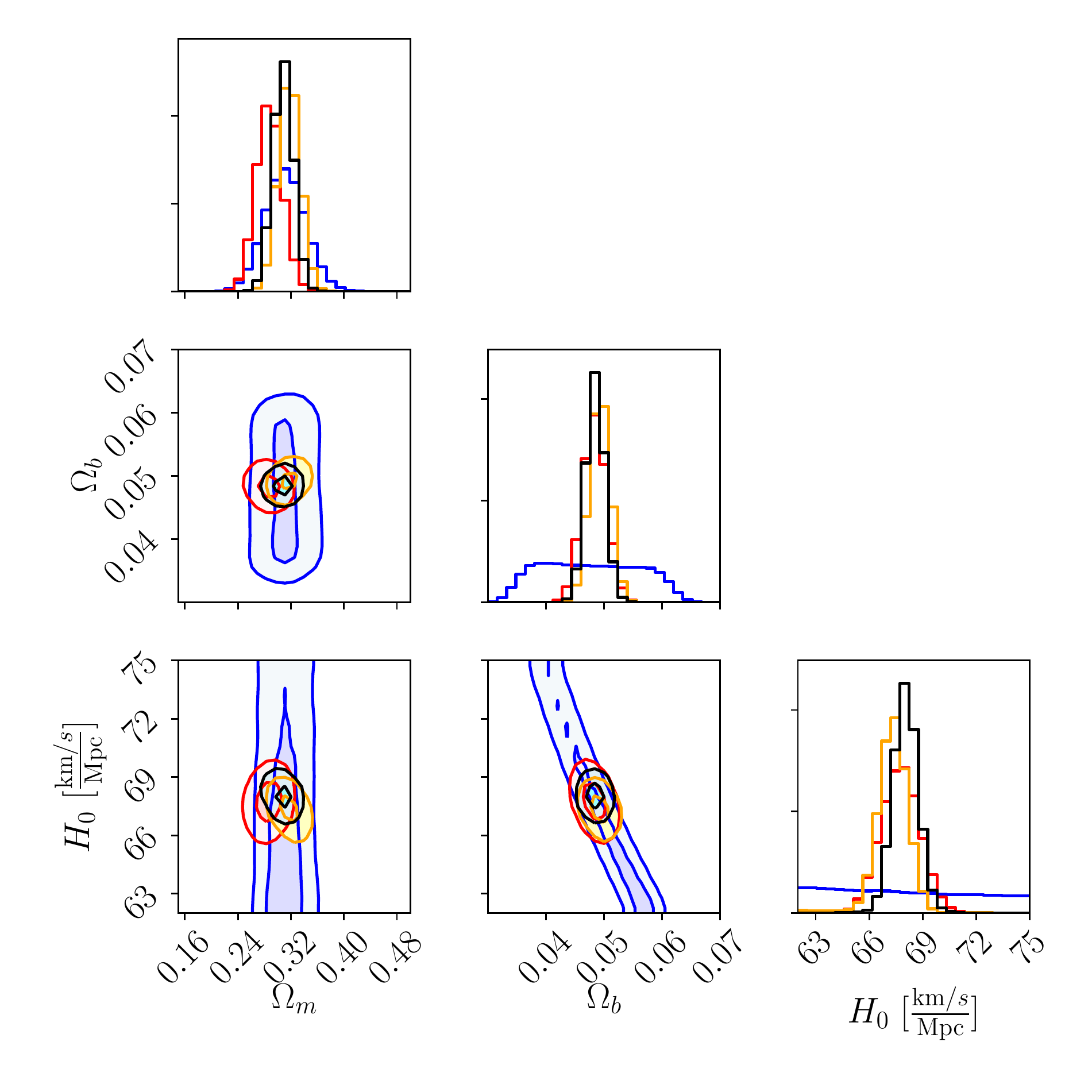}
    \subcaption{Posterior distribution in \LCDM}
    \label{fig:LCDMb}
    \end{subfigure}
    \caption{Results in \LCDM. \textit{Left:} Hubble diagram of the combined fit using all available data sets. The BIC is given in absolute numbers. \textit{Right}: Posterior distribution of model parameters with marginalised auxiliary parameters (including BBN prior). SN+quasars (blue), BAO (red), CMB (orange), all combined~(black). The contours represent $1\sigma$ and $2\sigma$ intervals.}
    \label{fig:LCDM}
\end{figure}

Marginalising over the nuisance parameters yields the compressed results displayed in Fig.~\ref{fig:LCDM} and summarised in Tab.~\ref{tab:results}. Here, we would like to note that our results agree well with the findings by the Planck collaboration, which find a slightly lower value for $H_0 = (67.4\pm 0.5) \frac{\mathrm{km}/\mathrm{s}}{\mathrm{Mpc}} $ ~\cite{Aghanim:2018eyx} (we find $H_0 = 68.4 \pm 0.4\frac{\mathrm{km}/\mathrm{s}}{\mathrm{Mpc}} $). The inclusion of more data sets does not affect the value of $H_0$ significantly and thus does not alleviate or worsen the long-standing tension between local calibrations of SNe (see e.g.~Refs.~\cite{Riess:2019cxk} for the most recent analysis) and the results from CMB measurements. In conclusion, we find that SNe and quasars yield compatible, tight constraints on $\Omega_m$, which is found to be $\Omega_m = 0.31\pm0.03$. 
Notice that neither SNe nor quasars can constrain $H_0$ alone because their absolute magnitudes are unknown [cf.~Eq.~\eqref{eq:sn_data}]. Via the inverse distance ladder approach, we break the degeneracy between $H_0$ and the magnitudes by including measurements of acoustic oscillation scale standard rulers. Combined, these allow us to constrain tightly the absolute magnitudes and $H_0$, as Fig.~\ref{fig:LCDMb} reveals.
In combination with the accurate determination of the location and height of the first CMB anisotropy peak, the parameters converge to $\Omega_m=0.308\pm0.006$ and $H_0 = (68.0\pm 0.4)\kmsmpc$, which is in good agreement with analyses of the CMB~\cite{Aghanim:2018eyx} with the results $H_0 = (67.4\pm 0.5)\kmsmpc$ and $\Omega_m=0.315\pm 0.007$, BAO~\cite{Aubourg:2014yra} with the results $H_0 = (67.3\pm 1.1)\kmsmpc$ and $\Omega_m=0.302\pm 0.008$, and the JLA SN sample~\cite{Betoule:2014frx} with result  $\Omega_m=0.295\pm 0.034$, establishing the robustness of our methodology.

\subsection{\LCDM\ with curvature}\label{sec:oLCDM}

\begin{figure}[t]
    \centering
    \hspace{-12mm}
    \begin{subfigure}[r]{.45\linewidth}
    \includegraphics[width = 1.1\textwidth]{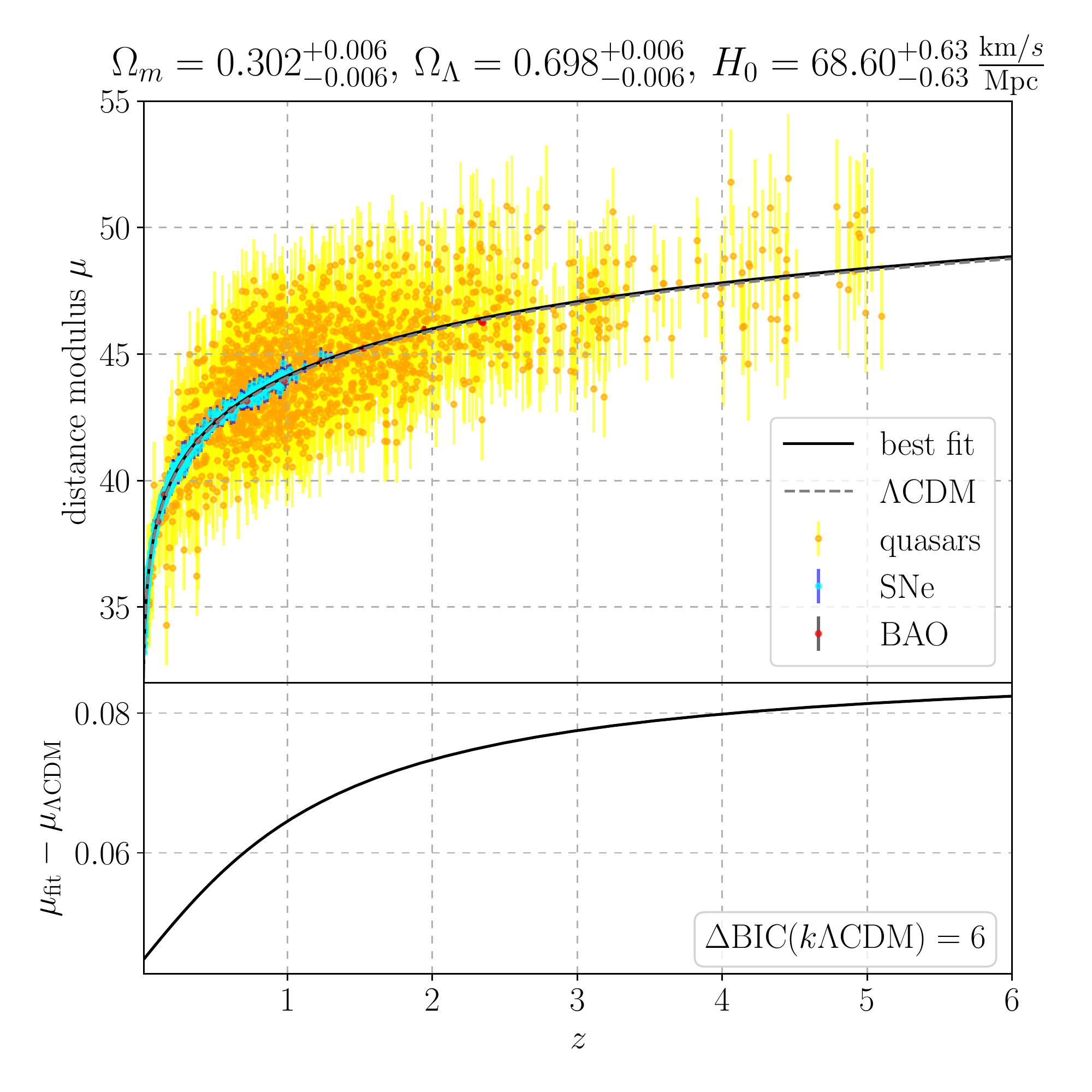}
    \subcaption{Hubble diagram in $k$\LCDM}
    \end{subfigure}
    \hspace{8mm}
    \begin{subfigure}[l]{.45\linewidth}
    \includegraphics[width = 1.1\textwidth]{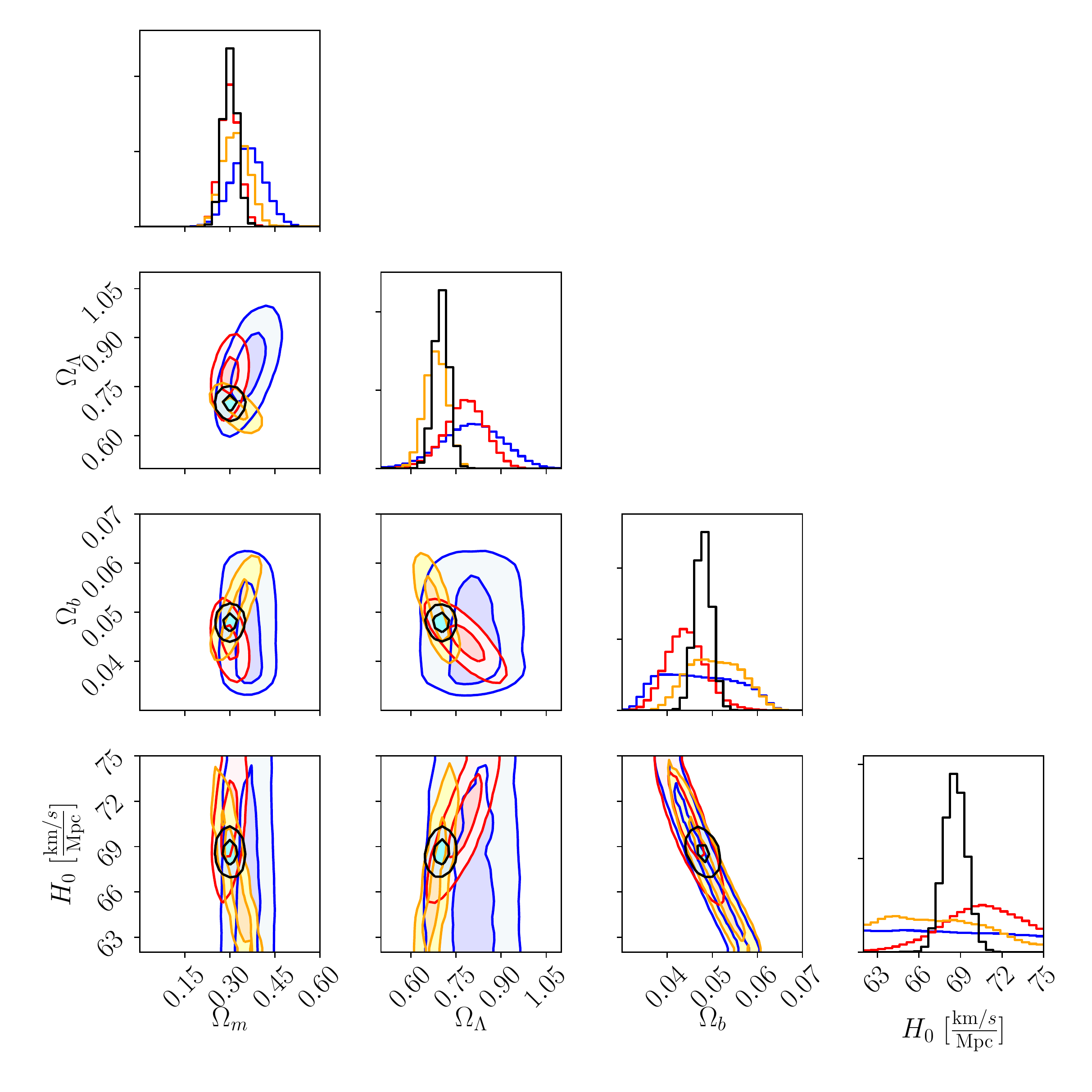}
    \subcaption{Posterior distribution in $k$\LCDM}
    \end{subfigure}
    \caption{Results in $k$\LCDM. \textit{Left:} Hubble diagram of the combined fit using all available data sets. \textit{Right}: Posterior distribution of model parameters with marginalised auxiliary parameters (including BBN prior). SN+quasars (blue), BAO (red), CMB (orange), combined (black). }
    \label{fig:oLCDM}
\end{figure}

Next, we modify the analysis carried out in the  previous section by relaxing the condition $\Omega_m+\Omega_\Lambda=1$ to include finite spatial curvature by introducing a new model parameter $\Omega_k = 1 - \Omega_m - \Omega_\Lambda$. A glance at the posterior distribution in Fig.~\ref{fig:oLCDM} and the results summarised in Tab.~\ref{tab:results} allows us to draw a number of interesting conclusions. 

It is conspicuous that the expected values for $\Omega_m$ and $\Omega_\Lambda$ are shifted to larger values once the quasar data is taken into account on top of the SN data, while in the previous case they were reasonably in accordance. This can be understood by recalling that quasars can be tested to much higher redshift so the effect of spatial curvature becomes relevant, effectively shifting $\Omega_m$ and $\Omega_\Lambda$ to much larger values compared to what is found for the SN data alone. 

One further interesting observation is that the inclusion of BAO and CMB data stabilises the values of the matter and dark energy densities at values close to the flat \LCDM\ case, $\Omega_m=0.302\pm0.006$, $\Omega_\Lambda = 0.698\pm0.006$ and $H_0 = 68.6\pm0.6\kmsmpc$. 

The curvature density is obtained from Eq.~\eqref{eq:oLCDM} and the results in Tab.~\ref{tab:results},
\begin{equation*}
    \Omega_k = 0.000 \pm 0.008 \,,
\end{equation*}
which hints at a flat universe. In fact, the statistical evidence shows that  $k$\LCDM\ is disfavoured with respect to \LCDM\ for any data sets considered in Tab.~\ref{tab:results}. Our results show that relaxing the flatness condition of \LCDM\ is not beneficial in terms of statistical evidence and the cosmological parameters converge nevertheless to a flat \LCDM\ universe, while adding another free parameter is penalised by the BIC.

\subsection{$w$\LCDM}\label{sec:wLCDM_results}

We find similar results if we parametrically extend the standard cosmology \LCDM\ to leave the equation of state parameter $w$ of dark energy an undetermined parameter. In this case, $w$\LCDM\ is strongly disfavoured compared to \LCDM\ for SN, SN+Q and SN+Q+BAO+CMB, and disfavoured for SN+Q+BAO. 
The curvature density here is found to be $\Omega_k=-0.001\pm0.013$ and the equation of state parameter turns out to be $w=-1.011\pm0.05$, values which are again very close to those of \LCDM.

\begin{figure}[t]
    \centering
    \hspace{-12mm}
    \begin{subfigure}[r]{.45\linewidth}
    \includegraphics[width = 1.1\textwidth]{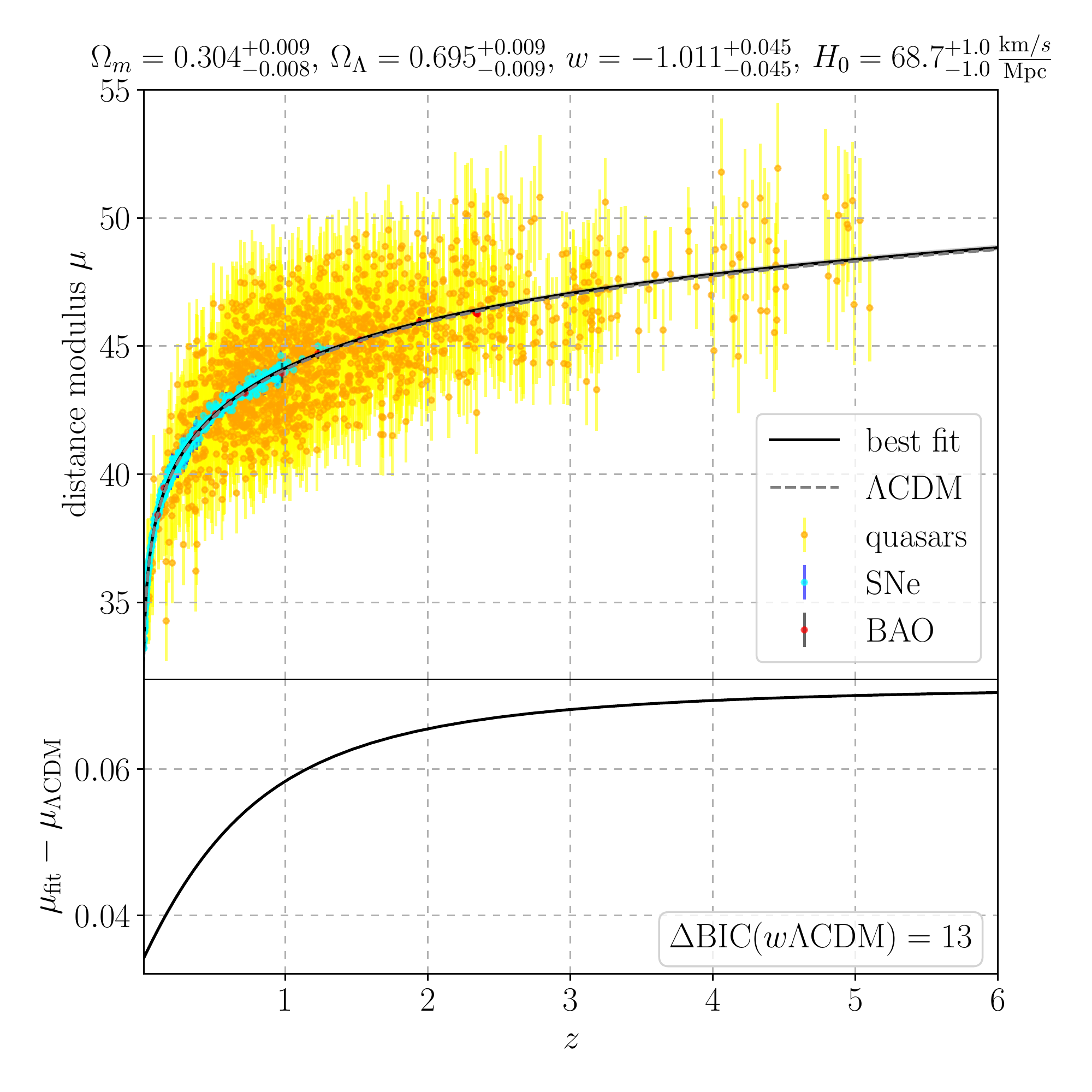}
    \subcaption{Hubble diagram in $w$\LCDM}
    \end{subfigure}
    \hspace{8mm}
    \begin{subfigure}[l]{.45\linewidth}
    \includegraphics[width = 1.1\textwidth]{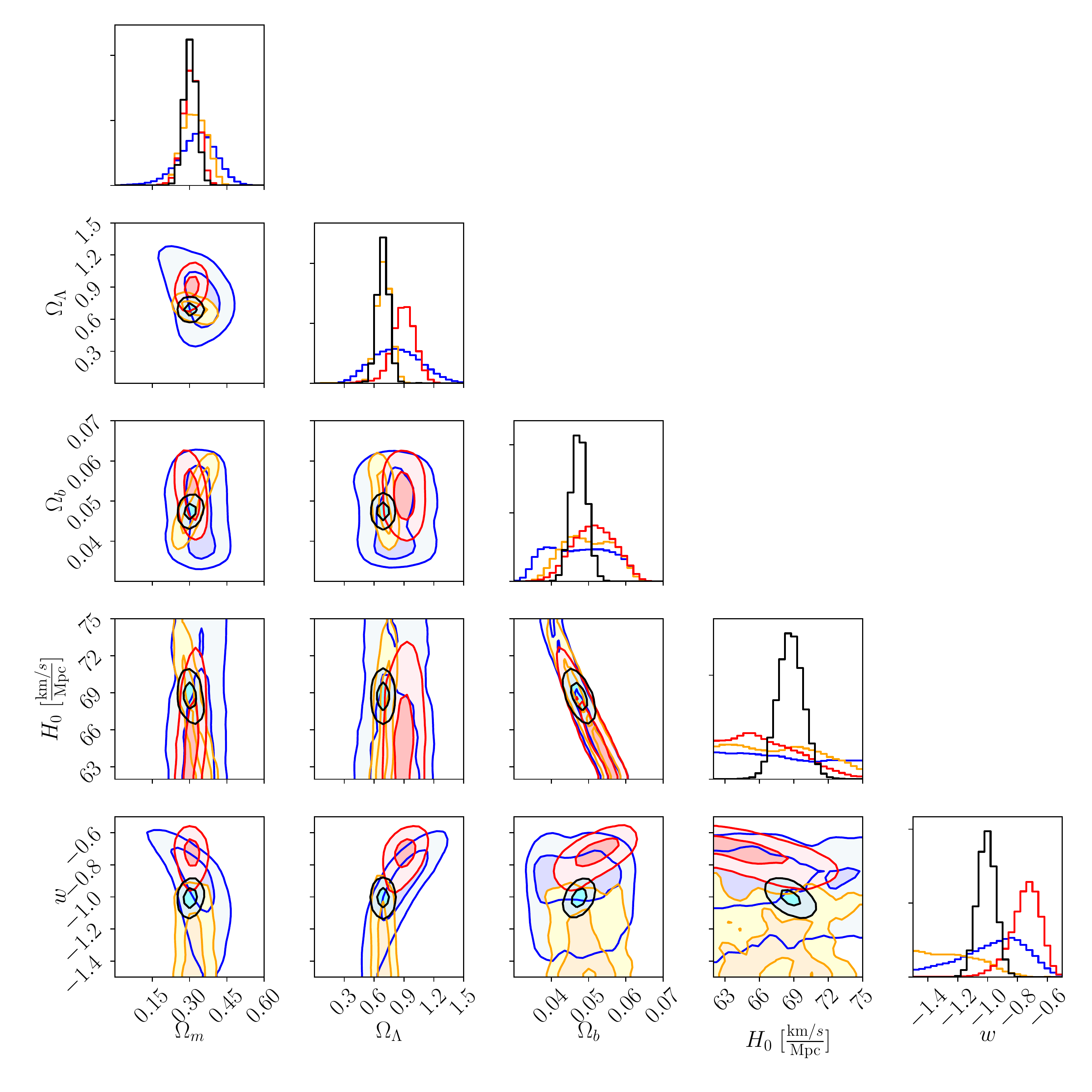}
    \subcaption{Posterior distribution in $w$\LCDM}
    \label{fig:wLCDMposterior}
    \end{subfigure}
    \caption{Results in $w$\LCDM. \textit{Left:} Hubble diagram of the combined fit using all available data sets. \textit{Right}: Posterior distribution of model parameters with marginalised auxiliary parameters (including BBN prior). SN+quasars (blue), BAO (red), CMB (orange), combined (black). }
    \label{fig:wLCDM}
\end{figure}

From Fig.~\ref{fig:wLCDM} we conclude that the additional parameters open up new regions of parameter space and intricate degeneracies arise, see e.g.~in the $w$--$\Omega_m$ marginalised posterior. Consequently, we see that certain data sets, e.g.~the BAO-only posterior, favour a much lower Hubble rate around $H_0 = 67.6 \kmsmpc$ compared to the combined fit. The reason lies in the fact that the CMB data do not allow an equation of state parameters much larger than $w=-1$ (as preferred by the BAO data), and the two parameters share precisely such a degeneracy, cf.~bottom left panel of Fig.~\ref{fig:wLCDMposterior}.

\newpage

\subsection{Bigravity cosmology}\label{sec:Bigra_results}

We now discuss the bigravity model fits for our setup.\footnote{Note that there are several possible modifications to our standard bigravity setup, which may considerably alter the cosmology. One such case is the choice of the matter coupling; e.g., see the analyses in Refs.~\cite{Enander:2014xga,Akrami:2018yjz} for bigravity with doubly coupled matter.} As shown in Sec.~\ref{sec:bigra_cosmology}, the parameters $m$ and $\alpha$ have been absorbed into the definition of the $B_i$, which leaves us with three bigravity fit parameters (on top of the usual number of parameters for \LCDM~and $k$\LCDM). This is the correct number of independent, physical bigravity parameters~\cite{Luben:2020xll}. Previous works on cosmological fits of bigravity include Refs.~\cite{Solomon:2014dua,Akrami:2012vf,Mortsell:2018mfj}. We improve on these results by the inclusion of the new quasar data set, and by performing the full Bayesian analysis over all bigravity parameters. Previous analyses have often restricted their attention to subsets of models, in which only one or two of the $B_i$ are non-zero (for a counterexample, see Ref.~\cite{Akrami:2012vf}). The main reason for restricting only to subsets of the parameter space in some previous work has been the fact that increasing the number of free parameters enlarges the volume of the parameter space, which then results in a larger Bayes factor. At the present time, we see no a~priori physical motif to restrict the bigravity parameter space, and have therefore only analysed the unrestricted model, i.e.~we fit \textit{all} bigravity parameters  $(B_{1,2,3})$ as well as a free parameter to describe curvature. The analysis in~\cite{Akrami:2012vf} has shown that the quality of the fit does not improve significantly when using a restricted model; however, due to the large number of possible correlations between the full set of bigravity parameters, a fit using a constrained model offers more insight into the structure of the bimetric model.

Nevertheless, the bigravity parameters are \textit{a~priori} allowed to take on any value in $\mathbb{R}$, and with the restriction on $B_1$ presented in Sec.~\ref{sec:bigra_cosmology}, the allowed ranges are $  B_1 > 0$ and $ B_2,B_3 \in \mathbb{R}$.

This parameter space could be sensibly scanned by a logarithmic sampling over a large range. However, it is reasonable to assume that the best fit cosmology closely resembles \LCDM. We therefore anticipate that the dynamical CC takes the current value $\Lambda(z=0) \sim (H_0^\text{\LCDM})^2$. Under the constraint of the master equation Eq.~\eqref{eq:bigra_master_eq}, this implies that all $B_i$ are of the same order, barring any fine tuning.
For the analysis, we therefore sample the bigravity parameters in the ranges $B_1 = [0,+100]$ and $B_{2,3} = [-100,+100]$ using flat priors. The chosen range ensures that the dynamical effects of bigravity are small, yet non-negligible: as Eq.~\eqref{eq:bigra_master_eq} shows, choosing a larger value for any of the $B_i$ decreases the redshift-dependence of the ratio of scale factors $y$, as the time-varying energy densities are suppressed in this case. However, with a constant $y$ the model's dynamics reproduce \LCDM.

\begin{figure}[t]
    \centering
    \hspace{-12mm}
    \begin{subfigure}[b]{.4\linewidth}
    \includegraphics[width = 1.1\textwidth]{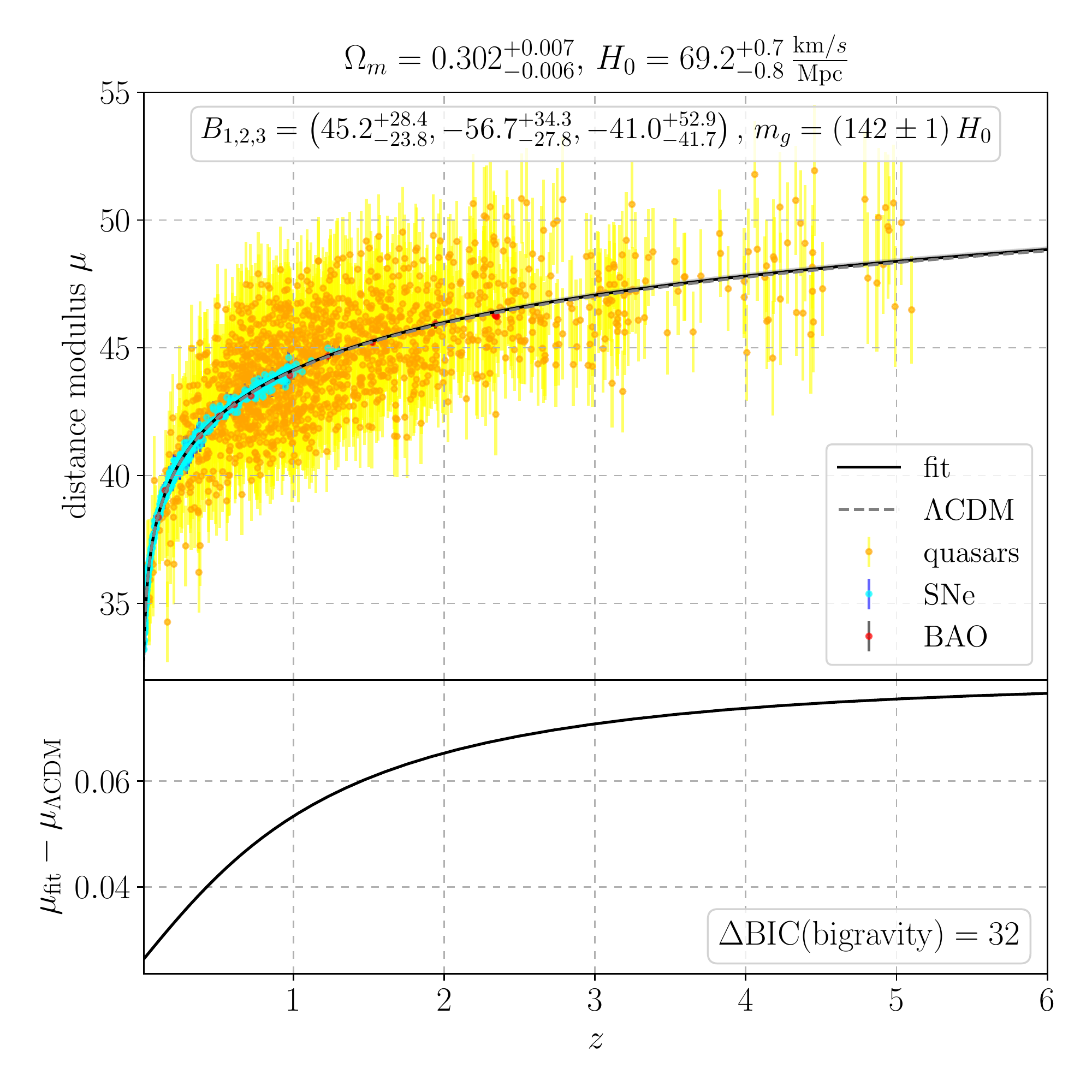}
    \subcaption{Hubble diagram in bigravity}
    \end{subfigure}
    \hspace{1mm}
    \begin{subfigure}[b]{.55\linewidth}
    \includegraphics[width = 1.1\textwidth]{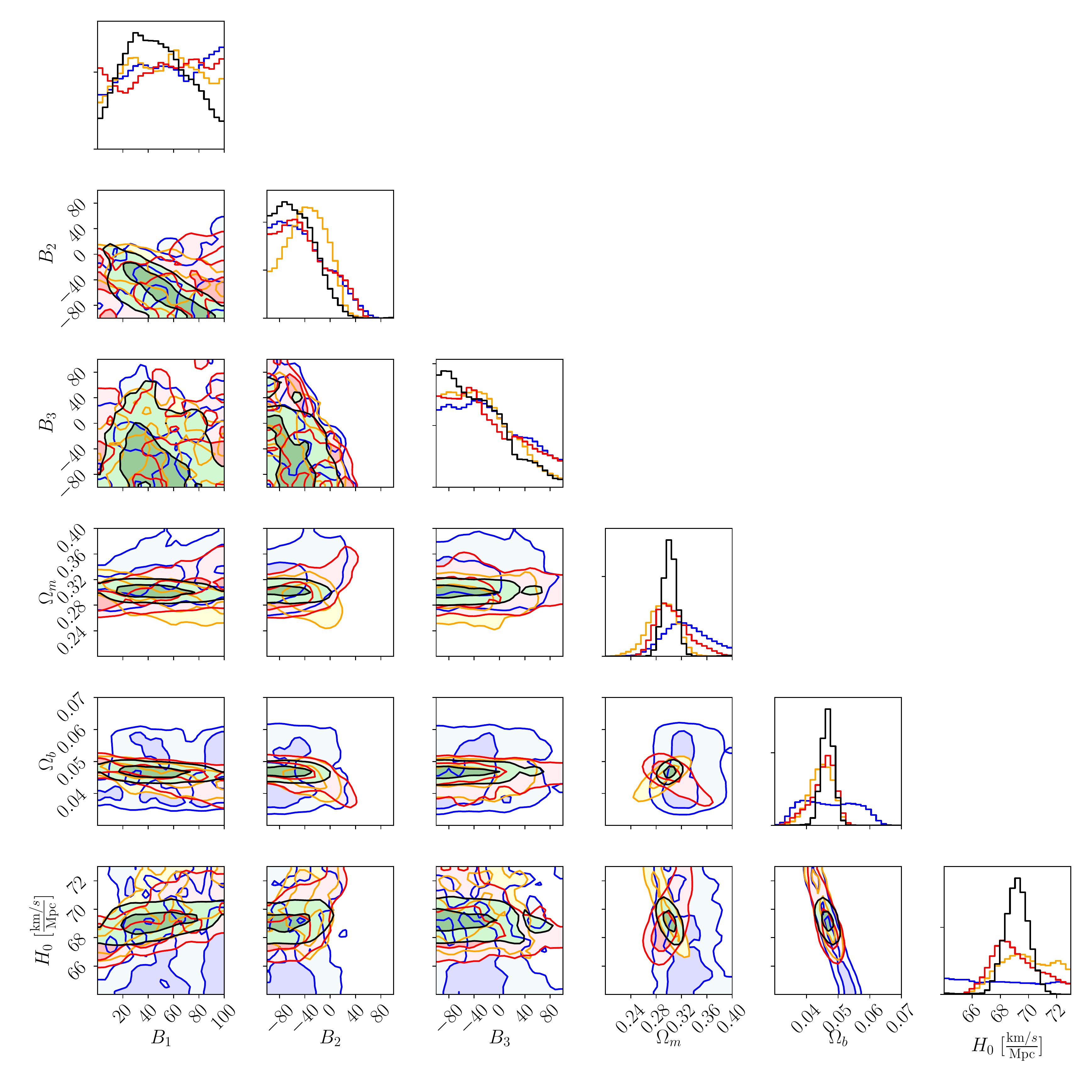}
    \subcaption{Posterior distribution in bigravity}
    \end{subfigure}
    \caption{Results in bigravity with flat geometry. \textit{Left:} Hubble diagram of the combined fit using all available cosmological data sets. The inset shows the best fit values of $B_{1,2,3}$ and the physical graviton mass as given by Eq.~\eqref{eq:bigra_graviton_mass_mg}. \textit{Right}: Posterior distribution of model parameters with marginalised auxiliary parameters (including BBN prior). SN+quasars (blue), BAO (red), CMB (orange), combined (black). }
    \label{fig:Bigra}
\end{figure}

\begin{figure}[h]
    \centering
    \hspace{-12mm}
    \begin{subfigure}[r]{.46\linewidth}
    \includegraphics[width = 1.05\textwidth]{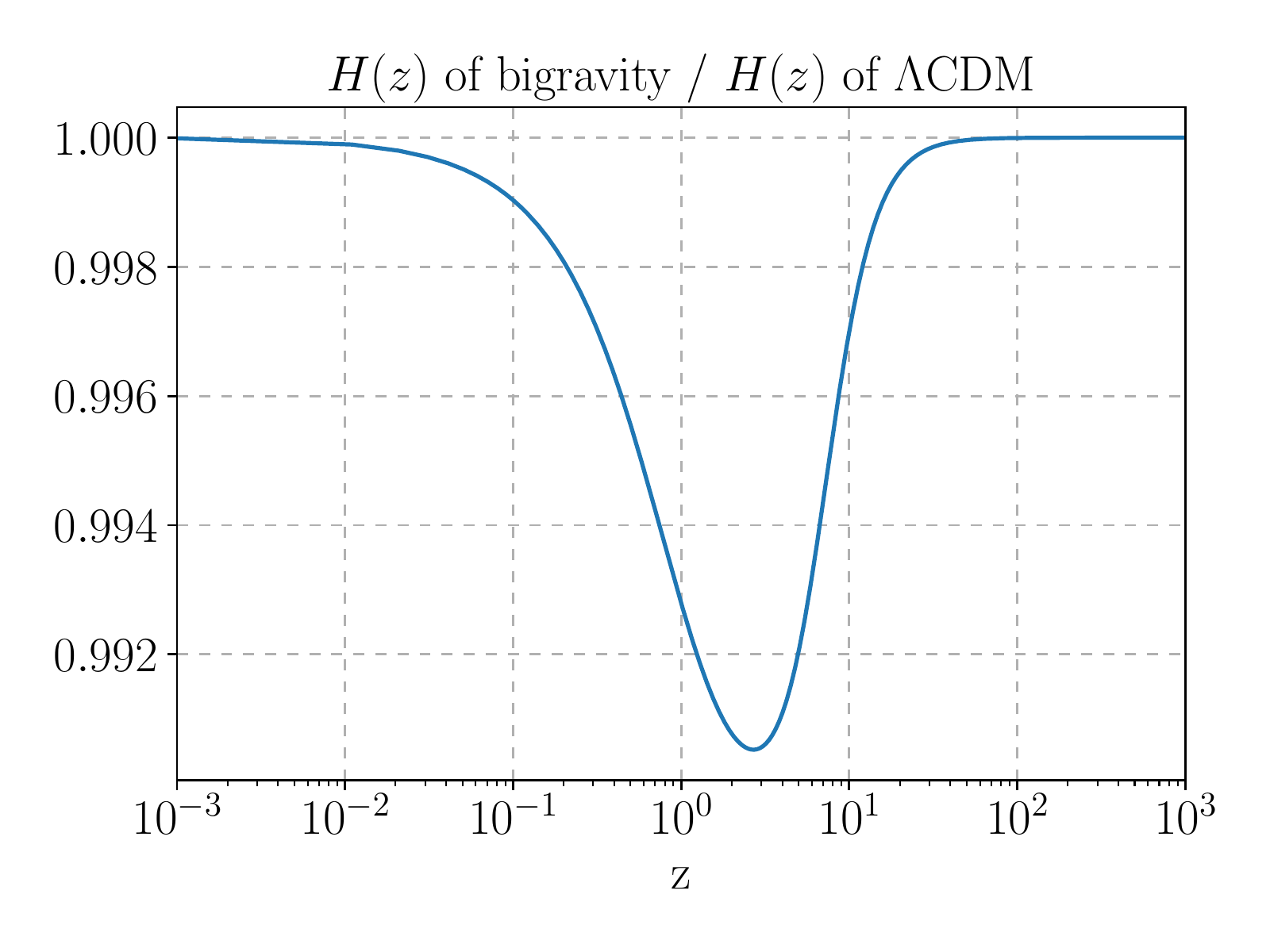}
    \subcaption{Hubble rate in bigravity over \LCDM}
    \end{subfigure}
    \hspace{8mm}
    \begin{subfigure}[l]{.46\linewidth}
    \includegraphics[width = 1.05\textwidth]{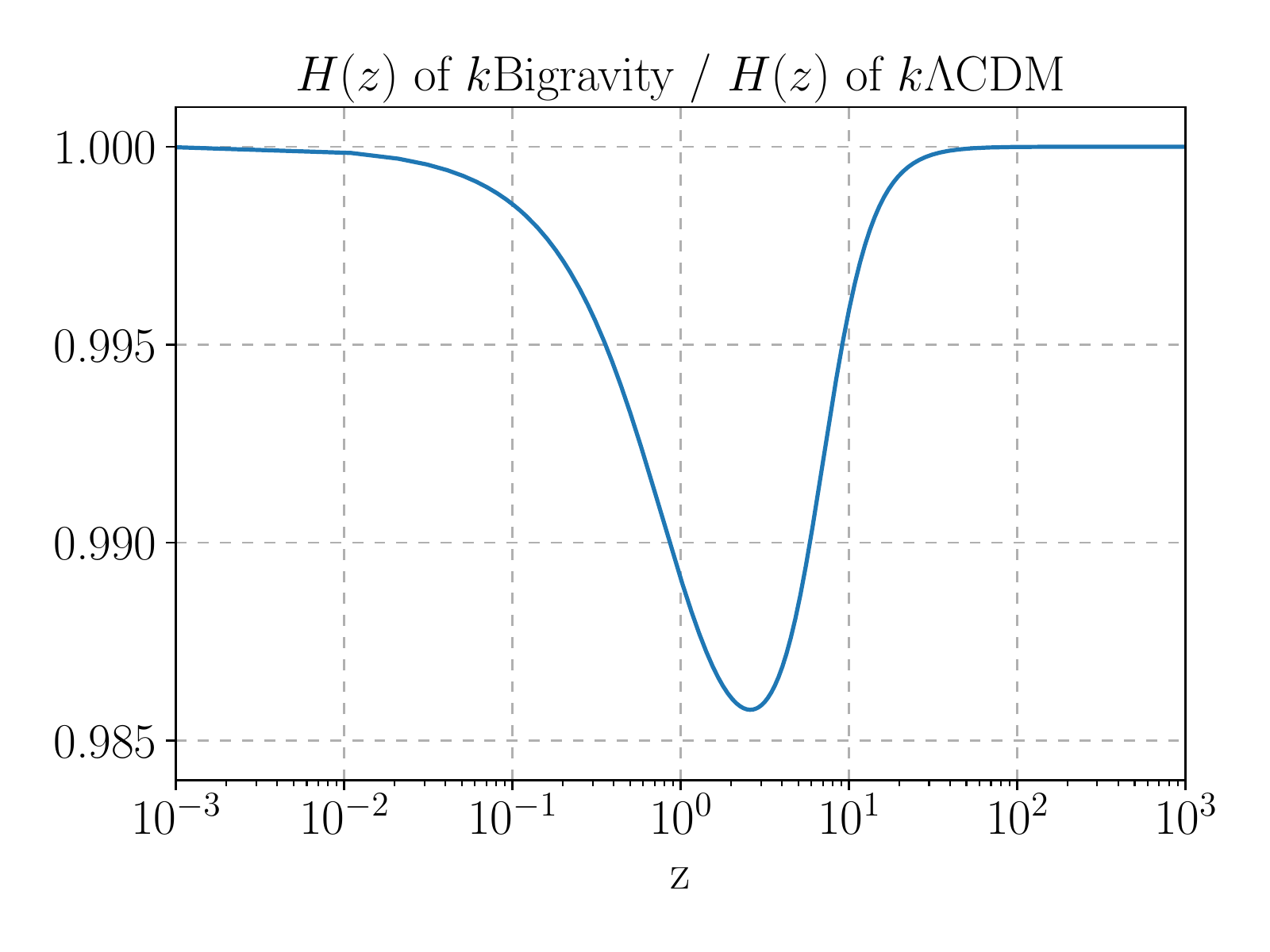}
    \subcaption{Hubble rate in $k$Bigravity over $k$\LCDM}
    \end{subfigure}
    \caption{Ratio of Hubble parameter $H(z)$ in ($k$)bigravity over ($k$)\LCDM\,. The cosmological parameters have been set to the best fit (considering all data sets) of the respective bimetric theory. As expected, the bimetric theories asymptotically match $(k)$\LCDM~at high and low~$z$.}
    \label{fig:Bigra_ratio_H}
\end{figure}

Fig.~\ref{fig:Bigra} shows the results in bigravity with zero curvature, where we find that the best fit values of $\Omega_m$, $\Omega_b$ and $H_0$ for the combined analysis depart only slightly from their counterparts in concordance cosmology and its related theories. For the bigravity parameters $B_{1,2,3}$, we find neither a clear sign of convergence, nor a simply visualised correlation between any pair of parameters.\footnote{As a check, we have also performed the fit using the logarithmic sampling variables $\log_{10}{\pm B_i}$, the results of which support these conclusions.} This can be understood by inspecting Fig.~\ref{fig:Bigra_ratio_H}: due to our choice of priors, the bigravity models which are probed by our setup deviate from \LCDM, however only at the percent level and in a redshift interval where the data is less constraining. In consequence, the standard cosmological parameters $(\Omega_m,\Omega_b, H_0)$ are well constrained, which also holds for the parameter which characterises the onset of the deviation from \LCDM, the graviton mass [Eq.~\eqref{eq:bigra_graviton_mass_mg}].
Using the best fit cosmology, we obtain $m_g~=~(142 \pm 1) \, H_0$, which complies with our expectation: it is a value close enough to $H_0$ so that the physical interpretation is that the graviton mass provides the dark energy component, but also satisfies the Higuchi ghost bound. It is also compatible with local tests of gravity and massive spin-2 states~\cite{Max:2017flc,Platscher:2018voh,Luben:2018ekw}.

While these results show that a consistent bigravity cosmology can be formulated, and that it is compatible with our range of observational tests, the pressing question is whether bigravity improves the fit compared to \LCDM. The value $\Delta\mathrm{BIC}=32$ reveals that this model is not preferred w.r.t.~simpler modifications of \LCDM. This is explained as this model mimics \LCDM~with zero curvature at large $z$, and thus is unable to improve the fit of the precisely known data sets BAO and CMB; but at the same time, the model brings with it an increased number of model parameters, which increases the BIC.

\begin{figure}
    \centering
    \hspace{-12mm}
    \begin{subfigure}[c]{.4\linewidth}
    \includegraphics[width = 1.1\textwidth]{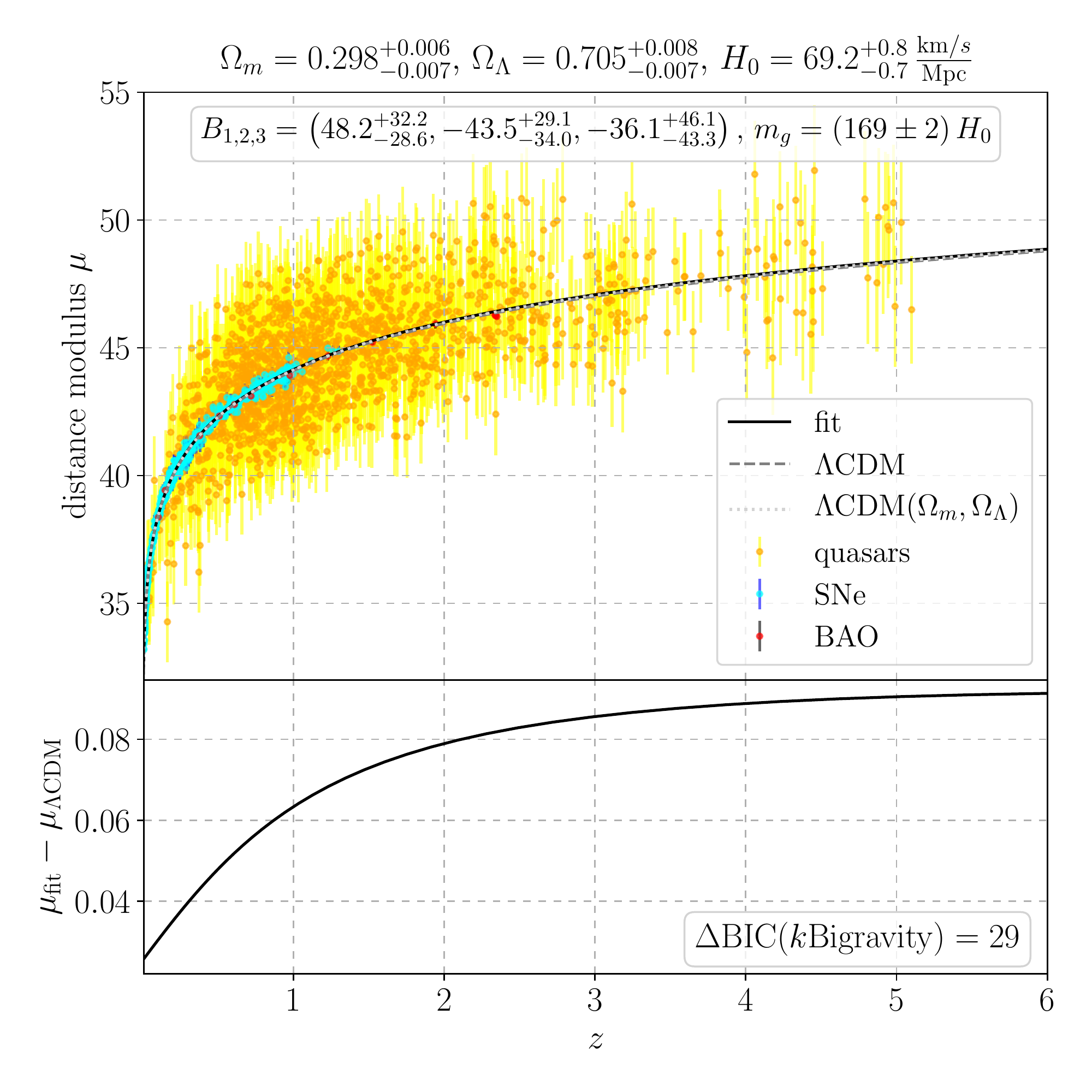}
    \subcaption{Hubble diagram in $k$Bigravity}
    \end{subfigure}
    \begin{subfigure}[l]{.8\linewidth}
    \includegraphics[width = 1.1\textwidth]{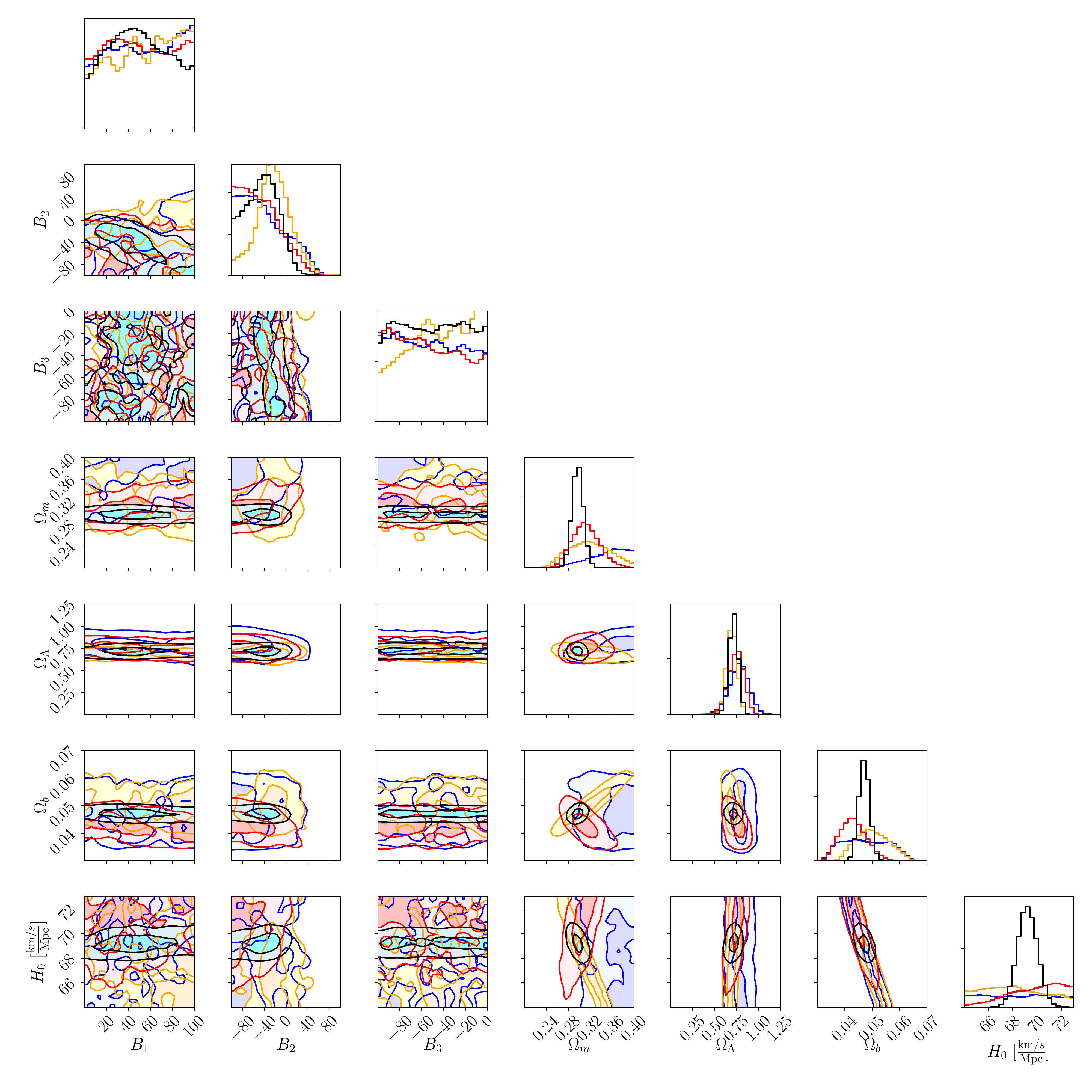}
    \subcaption{Posterior distribution in $k$Bigravity}
    \end{subfigure}
    \caption{Results in bigravity with curvature. \textit{Left:} Hubble diagram of the combined fit using all available cosmological data sets. The best fit values of $B_{1,2,3}$ and the physical graviton mass are shown as an inset. \textit{Right}: Posterior distribution of model parameters with marginalised auxiliary parameters (including BBN prior). SN+quasars (blue), BAO (red), CMB (orange), combined (black). }
    \label{fig:oBigra}
\end{figure}

With flat bigravity being strongly disfavoured, we thus turn to bigravity with a free curvature parameter in Fig.~\ref{fig:oBigra}. Again, we find all bigravity parameters to be $\order{100}$, while the mass density parameters and $H_0$ are similar to those found with~$(k)$\LCDM. The physical graviton mass is of the same order, $m_g~=~(169 \pm 2) \, H_0$. The curvature density is compatible with a flat universe. With these results it is not a surprise that we find $\Delta\mathrm{BIC} = 29$, indicating that bigravity with curvature is also strongly disfavoured, as it does not provide a significant improvement on the cosmological fit. However, one must keep in mind that these results do not rule out the possibility of a bimetric cosmology compatible with observations; the BIC is merely a statement about the improvement of the fit, while penalising the introduction of additional variables. From a model building point of view, bigravity still retains its desired features.

Furthermore, we comment on the fate of bimetric cosmologies at high $z$: as discussed in Sec.~\ref{sec:bigra_cosmology}, our chosen branch of bigravity must reduce to \LCDM~in this limit, as the second metric is effectively turned off when $y \to 0$. We have verified this behaviour numerically for the best fit parameters, see Fig.~\ref{fig:Bigra_ratio_H}. This shows that the best fit bigravity cosmologies match ($k$)\LCDM~at early times, i.e.~at redshifts upwards of $z=\,$10\,--\,100. Therefore, a fit involving only BAO and CMB measurements must yield the same result for ($k$)bigravity or ($k$)\LCDM; we have verified this as well. Following the argumentation of Ref.~\cite{Luben:2019yyx}, one can interpret this as the Vainshtein screening kicking in for large densities (large redshifts).

Finally, we discuss our choice of sampling the parameters $B_{1,2,3}$, fixing the redundant parameter $\alpha$. As we have seen, the results do not converge on the interval we are probing. However, we will now recall the discussion of perturbative stability at early times. In this light, our results should be interpreted as implementing the choice $\alpha = 1$. As shown in Ref.~\cite{Akrami:2015qga}, for $\alpha \ll 1$ and the $B_i$ within the same order, the instability of linear perturbartions can be shifted to very early times. By the rescaling invariance of the combination of parameters $\alpha^{-i}B_i$, an equivalent cosmology is given by the choice $\alpha =1$ while $B_1 \ll B_2 \ll B_3$. Clearly, for e.g.~$\alpha = 10^{-17}$, this is not accommodated by our choice of priors. However, we have also implemented a search for solutions of this type through logarithmic sampling; we do not find a statistically relevant cosmological fit for such a choice of parameters. This is easily understood when looking at the master equation~Eq.~\eqref{eq:bigra_master_eq}: for a large hierarchy between the $B_i$, these parameters need to be tuned to a high degree in order to allow for a dynamical solution of the master equation.

\newpage

\subsection{Conformal Gravity}\label{sec:CG_results}
Previous works on cosmological fits in CG can be found in Refs.~\cite{Mannheim:2005bfa,Diaferio:2011kc, Yang:2013skk, Roberts:2017nkm}. The first of these references uses SN data as standard candles; Ref.~\cite{Diaferio:2011kc} uses SNe and GRBs as standard candles; Ref.~\cite{Yang:2013skk} uses supernovae as standard candles and $H(z)$ measurements. In Ref.~\cite{Roberts:2017nkm} the model parameters are fixed to the best fit values of Ref.~\cite{Mannheim:2005bfa} and extrapolated to GRB and quasar data to account for the statistical evidence of these model parameters.

In the present work, we utilise the SNe and quasars data sets for an up-to-date assessment of the viability of CG cosmology compared to the base \LCDM~model. Note that we do not include the CMB measurements in this section, since the Planck analysis is based on a flat \LCDM~cosmology, and CG cosmology does not reduce to the concordance model at high redshift. For similar reasons, we exclude also the BAO data set. In this case, a careful treatment of the calculations of the drag epoch $z_d$ and the comoving sound horizon $c_s(z)$ (cf. Tab.~\ref{tab:BAOdata} in the Appendix) are required, a task which is beyond the scope of this work. Note that in this fit, the Hubble parameter remains unconstrained, as our SN and quasar samples are not calibrated in absolute magnitude.

\begin{figure}[t]
    \centering
    \hspace{-12mm}
    \begin{subfigure}[r]{.45\linewidth}
    \includegraphics[width = 1.1\textwidth]{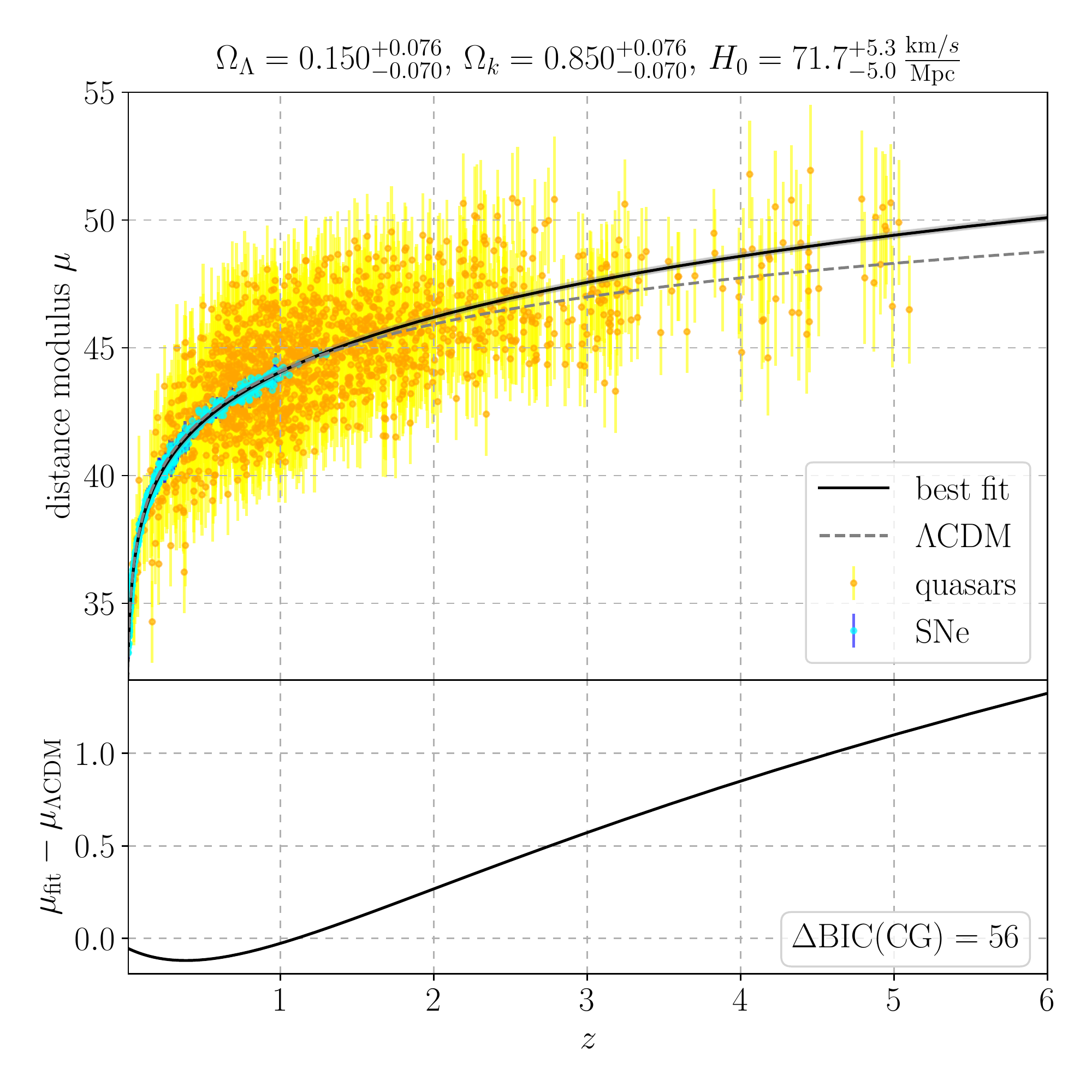}
    \subcaption{Hubble diagram in CG}
    \label{fig:CGa}
    \end{subfigure}
    \hspace{8mm}
    \begin{subfigure}[l]{.45\linewidth}
    \includegraphics[width = 1.1\textwidth]{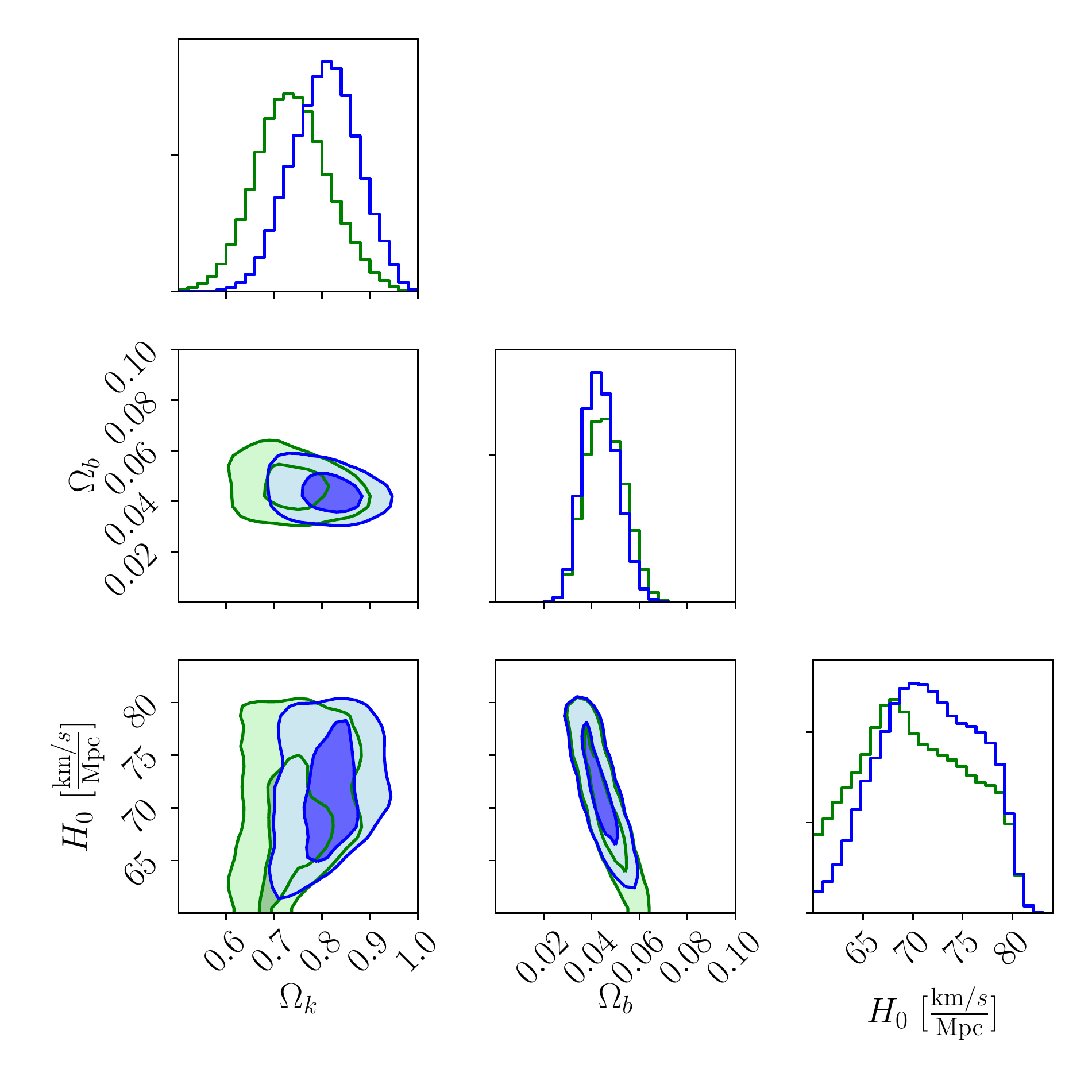}
    \subcaption{Posterior distribution in CG}
    \label{fig:CGb}
    \end{subfigure}
    \caption{Results in CG. \textit{Left:} Hubble diagram of the combined fit using SN and quasar data sets. \textit{Right}: Posterior distribution of model parameters with marginalised auxiliary parameters (including BBN prior). SN (green), SN+quasars (blue).}
    \label{fig:CG}
\end{figure}

As explained in Sec.~\ref{sec:CG}, we use Eq.~\eqref{eq:CGlowz} as the Friedmann equation valid for low redshifts. Hence, we choose $\Omega_k = 1 - \overline \Omega_\Lambda$ as free model parameter which can be tested by the SN+Q data set. The results are presented in Fig.~\ref{fig:CG} (see also Table \ref{tab:results}). Under consideration of only SN data, CG is disfavoured with respect to $\Lambda$CDM and becomes strongly disfavoured if quasars are included in the analysis.
The best fit value for the joint analysis of SNe and quasar data is $\Omega_k =  0.850^{+0.070}_{-0.081} $. This value for $\Omega_k$ agrees well with the results of Ref.~\cite{Diaferio:2011kc} which find $\Omega_k = 0.836 ^{+0.015}_{-0.022}$. However, Refs.~\cite{Mannheim:2005bfa, Yang:2013skk} find smaller values, $\Omega_k \approx 0.63$ and $\Omega_k = 0.67 \pm 0.06$, respectively. These deviations may be caused by the difference in the data sets which are considered. In particular, the observational data considered in Ref.~\cite{Diaferio:2011kc} and in this work reaches out to higher redshifts $z \sim 6$ compared to the data considered in Refs.~\cite{Mannheim:2005bfa, Yang:2013skk}.

Under the above considerations we are led to the conclusion that the cosmological model obtained from CG as outlined in Sec.~\ref{sec:CG} is strongly disfavoured with respect to the baseline \LCDM~cosmology. 

As we have outlined in Sec.~\ref{sec:CG}, CG has the unique feature that $\Omega_k$ can also be tested by  galactic dynamics and we find that this impairs the viability of CG further. To be more precise, the relation in Eq.~\eqref{eq:gammakrelation} enables us to independently infer the value of $\Omega_k$ from galactic rotation curves. For instance, the result of Ref.~\cite{OBrien:2017bwr} is
\begin{equation*}
    \gamma_0 = 3.06 \cdot 10^{-30} \text{cm}^{-1} \quad \Rightarrow \quad \Omega_k = 4.12 \cdot 10^{-4} \,,
\end{equation*}
if it is assumed that CG addresses the missing mass problem of galaxies without invoking a dark matter component. This result was obtained from a fit of galactic rotation curves of 207 galaxies. A severe tension of the above value for $\Omega_k$ with our results obtained from cosmological data is manifest, cf.~Tab.~\ref{tab:results}. It is clear that the observed galactic dynamics demand a significant smaller value of $\Omega_k$ (or equivalently $\gamma_0$) than the observations on cosmological scales which we consider in this work. Hence, if it is assumed that CG explains galactic rotation curves without dark matter, a consistent reconciliation of both phenomena seems unlikely. In fact, we have performed a joint analysis of cosmological and galactic data which is based on the relation in Eq.~\eqref{eq:gammakrelation} and we did not find sensible results. Furthermore, we can confirm that we find a similar value of $\gamma_0$ as in Ref.~\cite{Mannheim:2012qw} from our own analysis of the SPARC rotation curve data set. These considerations allow us to make the statement that CG is not able to address the missing mass problem without invoking dark matter and to account for a viable cosmological evolution simultaneously. However, with a dark matter component present as in \LCDM~one has to take only the constraints imposed by cosmology into account.

\begin{table}[p]
    \centering
    \resizebox{!}{11cm}{
    \begin{tabular}{r|c|cccc}
    \toprule
        model & parameter & SN & SN+Q & SN+Q+BAO & SN+Q+BAO+CMB\\
    \midrule
         \multirow{4}{*}{\LCDM}& $\Omega_m$ & $0.296^{+0.030}_{-0.028}$ & $0.311^{+0.028}_{-0.027}$ & $0.292^{+0.013}_{-0.012}$ & $0.3081^{+0.0063}_{-0.0057}$\\
         & $100\,\Omega_b$ & $4.71^{+0.93}_{-0.87}$ & $4.74^{+0.96}_{-0.87}$ & $4.82^{+0.13}_{-0.13}$ & $4.85^{+0.05}_{-0.05}$\\
         & $H_0\ [\kmsmpc]$ & $68.6^{+7.4}_{-6.0}$ & $68.4^{+7.3}_{-6.0}$ & $67.86^{+0.94}_{-0.93}$ & $68.04^{+0.43}_{-0.44}$\\
         & $\mathrm{BIC}$ & $-447$ & $2722$ & $2850$ & $2825$\\
    \midrule
         \multirow{5}{*}{$k$\LCDM}& $\Omega_m$ & $0.21^{+0.10}_{-0.10}$ & $0.36^{+0.049}_{-0.050}$ & $0.307^{+0.017}_{-0.017}$ & $0.302^{+0.006}_{-0.006}$\\
         & $\Omega_\Lambda$ & $0.55^{+0.19}_{-0.19}$ & $0.80^{+0.10}_{-0.10}$ & $0.775^{+0.051}_{-0.054}$ & $0.698^{+0.006}_{-0.006}$\\
         & $100\,\Omega_b$ & $4.77^{+0.93}_{-0.89}$ & $4.68^{+0.94}_{-0.84}$ & $4.39^{+0.36}_{-0.34}$ & $4.80^{+0.09}_{-0.09}$\\
         & $H_0\ [\kmsmpc]$ & $68.2^{+7.5}_{-5.8}$ & $68.9^{+7.2}_{-6.1}$ & $71.1^{+2.9}_{-2.8}$ & $68.60^{+0.63}_{-0.63}$\\
         & $\Delta\mathrm{BIC}$ & \textcolor{red}{$+6$} & \textcolor{red}{$+6$} & \textcolor{red}{$+6$} & \textcolor{red}{$+6$}\\
    \midrule
         \multirow{6}{*}{$w$\LCDM}& $\Omega_m$ &  $0.173^{+0.110}_{-0.096}$ & $0.335^{+0.064}_{-0.073}$ & $0.312^{+0.018}_{-0.018}$ & $0.304^{+0.009}_{-0.008}$\\
         & $\Omega_\Lambda$ & $0.45^{+0.35}_{-0.21}$ & $0.82^{+0.24}_{-0.23}$ & $0.899^{+0.069}_{-0.072}$ & $0.695^{+0.009}_{-0.009}$\\
         & $100\,\Omega_b$ & $4.68^{+0.94}_{-0.85}$ & $4.76^{+0.91}_{-0.91}$ & $4.61^{+0.42}_{-0.35}$ & $4.74^{+0.14}_{-0.14}$\\
         & $H_0\ [\kmsmpc]$ & $68.8^{+7.3}_{-6.0}$ & $68.3^{+7.6}_{-5.7}$ & $69.3^{+2.8}_{-2.9}$ & $68.73^{+0.97}_{-0.96}$\\
         & $w$ & $-1.09^{+0.32}_{-0.68}$ & $-0.96^{+0.20}_{-0.35}$ & $-0.821^{+0.054}_{-0.062}$ & $-1.011^{+0.045}_{-0.045}$\\
         & $\Delta\mathrm{BIC}$ & \textcolor{BrickRed}{$+13$} & \textcolor{BrickRed}{$+14$} & \textcolor{red}{$+7$} & \textcolor{BrickRed}{$+13$}\\
    \midrule[1.5pt]
         \multirow{9}{*}{Bigravity}& $\Omega_m$ & $0.319^{+0.059}_{-0.035}$ & $0.330^{+0.049}_{-0.034}$ & $0.305^{+0.025}_{-0.015}$ & $0.3016^{+0.0071}_{-0.0063}$\\
         & $100\,\Omega_b$ & $4.67^{+0.91}_{-0.82}$ & $4.7^{+0.95}_{-0.85}$ & $4.64^{+0.21}_{-0.39}$ & $4.66^{+0.11}_{-0.10}$\\
         & $H_0\ [\kmsmpc]$ & $68.9^{+7.0}_{-5.9}$ & $68.7^{+7.2}_{-6.1}$ & $69.2^{+3.1}_{-1.6}$ & $69.17^{+0.71}_{-0.78}$\\
         & $B_1$ & $53^{+32}_{-34}$ & $53^{+34}_{-33}$ & $53^{+32}_{-35}$ & $44^{+29}_{-23}$ \\
         & $B_2$ & $-40^{+49}_{-40}$ & $-48^{+52}_{-35}$ & $-53^{+47}_{-32}$ & $-55^{+35}_{-30}$\\
         & $B_3$ & $-21^{+73}_{-53}$ & $-24^{+70}_{-51}$ & $-24^{+73}_{-52}$ &  $-41^{+51}_{-40}$\\
         & $\Delta\mathrm{BIC}$ & \textcolor{BrickRed}{$+20$} & \textcolor{BrickRed}{$+23$} & \textcolor{BrickRed}{$+24$} & \textcolor{BrickRed}{$+32$}\\
     \midrule
      \multirow{10}{*}{$k$Bigravity}& $\Omega_m$ & $0.24^{+0.12}_{-0.11}$ & $0.386^{+0.069}_{-0.059}$ & $0.317^{+0.024}_{-0.020}$ & $0.2974^{+0.0068}_{-0.0071}$\\
      & $\Omega_\Lambda$ & $0.53^{+0.18}_{-0.17}$ & $0.780^{+0.092}_{-0.107}$ & $0.755^{+0.053}_{-0.058}$ & $0.7056^{+0.0079}_{-0.0076}$\\
      & $100\,\Omega_b$ & $4.84^{+0.87}_{-0.90}$ & $4.77^{+0.95}_{-0.89}$ & $4.24^{+0.39}_{-0.39}$ & $4.70^{+0.11}_{-0.11}$\\
      & $H_0\ [\kmsmpc]$ & $67.8^{+7.2}_{-5.5}$ & $68.3^{+7.3}_{-6.0}$ & $72.4^{+3.5}_{-3.2}$ & $69.23^{+0.77}_{-0.72}$\\
       & $B_1$ & $52^{+32}_{-32}$ & $53^{+34}_{-35}$ & $47^{+35}_{-33}$ & $49^{+32}_{-29}$\\
       & $B_2$ & $-43^{+50}_{-41}$ & $-46^{+53}_{-37}$ & $-58^{+45}_{-28}$ & $-41^{+29}_{-33}$\\
       & $B_3$ & $-18^{+73}_{-57}$ & $-25^{+76}_{-53}$ & $-24^{+74}_{-54}$ & $-36^{+45}_{-44}$\\
         & $\Delta\mathrm{BIC}$ & \textcolor{BrickRed}{$+26$} & \textcolor{BrickRed}{$+30$} & \textcolor{BrickRed}{$+30$} & \textcolor{BrickRed}{$+29$}\\
     \midrule[1.5pt]
        \multirow{5}{*}{CG}& $\Omega_k$ & $0.772^{+0.081}_{-0.068}$ & $0.850^{+0.070}_{-0.081}$ & --  & --\\
        & $100\,\Omega_b$ & $4.51^{+0.83}_{-0.73}$ & $4.32^{+0.67}_{-0.57}$ & -- & --\\
        & $H_0\ [\kmsmpc]$ & $70.0^{+6.6}_{-5.6}$ & $71.6^{+5.4}_{-4.9}$ &-- & --\\
        & $\Delta\mathrm{BIC}$ & \textcolor{red}{$+9$} & \textcolor{BrickRed}{$+56$} & --  & --  \\
    \bottomrule
    \end{tabular}
    }
    \caption{Summary of the results in the different model as discussed in Sec.~\ref{sec:results}. The BIC in \LCDM\ is given in absolute numbers, while all others are relative w.r.t~the \LCDM\ best fit. The color indicates the statistical significance: \textcolor{OliveGreen}{strong support} ($\Delta\mathrm{BIC}<-12$), \textcolor{green}{favourable} ($\Delta\mathrm{BIC}<-6$), \textcolor{orange}{inconclusive} ($\Delta\mathrm{BIC}<6$), \textcolor{red}{disfavoured} ($\Delta\mathrm{BIC}<12$), \textcolor{BrickRed}{strongly disfavoured} ($\Delta\mathrm{BIC}\ge12$) with respect to~\LCDM; see the Appendix for details.}
    \label{tab:results}
\end{table}

\clearpage

\section{Conclusions and outlook}
\label{sec:conclusions}
In this work we have performed a combined analysis of standard candles and standard rulers to account for the viability of six cosmological models: flat \LCDM, \LCDM\ with curvature, \LCDM\ with curvature and dynamical dark energy, bigravity, bigravity with curvature and conformal gravity (CG) cosmology. To this end, we have employed various data sets in the form of the joint light-curve analysis SN compilation, measurements of the BAO scale in the large scale structure, and the CMB measurement of the acoustic scale. In addition, we have extended this list for the first time by quasar measurements, which only recently have been proposed to serve as standard candles~\cite{Risaliti:2015zla,Risaliti:2018reu}. Although these measurements are afflicted with large uncertainties, they add many new standard candles at a previously unprobed range of high redshifts $1 \lesssim z \lesssim 6$ (complementary to the SN measurements at lower redshifts $z \lesssim 1$ and the CMB measurements at very high redshift). This enables us to test cosmological models on a wider range of scales, and thus to estimate cosmological parameters better. Recently, the same quasar data set has been utilized to test the \LCDM\ model and its parametric extensions in Ref.~\cite{Khadka:2019njj}, albeit not in conjunction with SN measurements. The analysis therein draws similar conclusions to the ones presented in this study.

Utilising the flat \LCDM~concordance model, we established the robustness of our methods by comparing our results to the literature. Considering our data sets, we have found in all cases that the modifications $k$\LCDM\ and $w$\LCDM\ are not favoured with respect to the concordance cosmology, with the latter even being strongly disfavoured. In both cases the deviation from a flat \LCDM\ universe is small, i.e.~close to flat and the equation of state is $w \approx -1$, if the complete data set is considered. Furthermore, the remaining cosmological parameters converge to values close to those found in \LCDM\ and no alleviation to the $H_0$ tension is present in these models.

Moving on to bigravity, our results show that the best fit cosmologies in this framework closely approximate \LCDM. The differentiation between bimetric theory and concordance cosmology is irrelevant at the time when CMB and BAO are set. At smaller redshift, where deviations from \LCDM\ are expected, bigravity is not able to improve the fit, and is strongly disfavoured from a statistical point of view -- irrespective of the geometry which is assumed (flat or with curvature).

Similarly, definitive conclusions can be drawn for CG cosmology. While the SN data alone already suggests that CG is disfavoured with respect to \LCDM, testing the model at higher redshifts with quasar measurements impairs the viability further. In addition, the curvature parameter we deduce from our results is in considerable tension with results from galactic surveys, if CG is also to account for the missing mass in galaxies without the addition of dark matter. This leaves no other conclusion than discarding this version of CG, where no dark matter in the universe is assumed, to describe both galactic and cosmic dynamics.

We hope that our results might hint to new avenues for cosmological model building based on modifications of GR. To this end, we give a transparent description of our methodology in the Appendix in conjunction with our code publicly available at~\cite{GitHub} including the aforementioned quasar data set.

We stress that our approach is solely focused on the level of the background cosmology, and that it would be desirable to extend this study to the computation of primordial temperature fluctuations. In this way, the cosmological models could be confronted with the measurement of the full CMB spectrum.

As far as CG is concerned, such an analysis is at present not available owing to the fact that it involves higher-order equations which render the computations much more involved than the standard \LCDM\ case. For the case of bigravity, we refer to the discussion in Sec.~\ref{sec:bigra_cosmology}. In summary, also in this case a full analysis of the cosmological perturbations remains as an open question.

\section*{Acknowledgements}
The authors would like to thank E.~Lusso for providing the quasar data set. J.~R. is supported by the IMPRS-PTFS. K.~M.~thanks the Dreiner group at University of Bonn for hospitality, during which part of this work was completed.

\appendix

\section{Data sets and analysis methods}
\label{app:data}
In this appendix we describe the data samples we have used in this and document our data analysis methods. The contents of this appendix will enable the inclined reader to reproduce all of our results. Furthermore, we provide the code to reproduce our results at~\cite{GitHub}.

\subsection{Distance measures}

Having specified a given model in terms of its Hubble rate's dependence on redshift, testable observables can be derived. To this end, objects of known brightness (standard candles) and known size (standard rulers) are identified at a certain redshift in order to infer the corresponding cosmic distance measures. We define the co-moving distance as a function of redshift
\begin{equation}
    d_C(z) = \int_0^z \frac{\D z'}{H(z')} ,
\end{equation}
from which a number of useful distance measures can be derived. 

\noindent\textbf{Standard candles} are objects of known brightness and their \emph{luminosity distance} is given by
\begin{equation}
    d_L(z) = (1+z) \Phi_k\left(d_C(z)\right)\,,
\end{equation}
with 
\begin{equation}
    \Phi_k(x) = 
    \begin{cases}
        \sqrt{\Omega_k H_0^2}^{-1} \sinh\left(\sqrt{\Omega_k H_0^2}\, x \right) ,& k< 0,\\
        x ,&k=0,\\
        \sqrt{\left|\Omega_k\right| H_0^2}^{-1} \sin\left(\sqrt{\left|\Omega_k\right|H_0^2}\, x \right), & k>0.
    \end{cases}
\end{equation}

\noindent\textbf{Standard rulers} are objects of known size, such as the BAO scale, and one measure their \emph{angular diameter distance},
\begin{equation}
    d_A(z) = \frac{\Phi_k\left(d_C(z)\right)}{1+z} = \frac{d_L(z)}{(1+z)^2}\,.
\end{equation}

\subsection{Big bang nucleosynthesis}

Our analysis assumes that BBN proceeds in the standard manner. In order to be in agreement with measurements of the primordial deuterium abundance, we combine \emph{all} likelihoods with a Gaussian prior on $100\,\Omega_b h^2= 2.22\pm 0.05$. This is the `conservative BBN prior' of Planck~2018~\cite{Aghanim:2018eyx} on the basis of the deuterium abundance measurement by Cooke~et~al.~\cite{Cooke:2017cwo}.

\subsection{Supernova data}
\label{subsec:SN_data}
In order to employ the power of SN standard candles, we make use of the Joint Light Curve Analysis (JLA) data~\cite{Betoule:2014frx}, a combined analysis of the available SDSS and SNLS data including very low ($z<0.1$) and high redshift data points ($z\gtrsim1$). The resulting set of 740 SN events, available from~\cite{JLA}, have previously been used to discriminate cosmological models, see e.g.~Ref.~\cite{Melia:2018kyb} for recent work. The distance modulus of a generic SN event is defined as $\mu = 5\, \log_{10}\left(\frac{d_L(z)}{1\Mpc}\right) + 25$, and can be related to the absolute and apparent bolometric magnitude of the given SN as,
\begin{equation}
\label{eq:sn_data}
    \mu = m_B + \alpha'\,X_1-\beta\,C-M_B\,,
\end{equation}
where $m_B$ and $M_B$ are apparent and absolute B-band magnitudes, respectively; $X_1$ characterises the shape of the SN light curve; and $C$ its deviation from the standard type Ia SN color. While $m_B$, $X_1$ and $C$ are measured, $\alpha'$, $\beta$ and $M_B$ need to be extracted from a \emph{joint fit} of the data to a given cosmological background model.\footnote{Taking $M_B$ as a fit parameter is the result of our ignorance about the absolute magnitude of the SN luminosity. Because this introduces an arbitrary rescaling of $d_L$, we are not able to infer $H_0$ from the SN fit alone. This is only possible if one includes a local calibrator (see~\cite{Riess:2019cxk}).} As proposed by the JLA analysis, we include an `adjustment parameter' $\Delta M_B$ for SNe in host galaxies with a masses $>10^{10}\Msol$, i.e.
\begin{equation}
    M_B =\begin{cases}M_B^0 & \text{if}\ M_\text{host} \le 10^{10}\Msol,\\ M_B^0 +\Delta M_B & \text{if}\ M_\text{host} > 10^{10}\Msol. \end{cases}
\end{equation}
This, together with a given model prediction for $d_L(z)$, allows us set up our log-likelihood for the SN data, most compactly written in matrix notation,
\begin{equation}\label{eq:SN_likelihood}
    -2 \log \mathcal{L}_\text{SN} = \left[ \vec{\mu} - \vec{\mu}_\text{model} \right]^T \mathbf{C}^{-1} \left[ \vec{\mu} - \vec{\mu}_\text{model} \right] + \log[\det (\mathbf{C})],\footnote{For practical reasons, the last term needs to be implemented as $\log[\det (\mathbf{C})] = \mathrm{tr}[\log(\mathbf{C})]$ as otherwise the determinant is below machine precision and this term evaluates to $-\infty$.}
\end{equation}
with the covariance matrix $\mathbf{C}$ decomposed into
\begin{equation}
    \mathbf{C} = \mathbf{D}_\text{stat} + \mathbf{C}_\text{stat} + \mathbf{C}_\text{sys}\,,
\end{equation}
and the diagonal matrix $\mathbf{D}_\text{stat}$ given as
\begin{equation}
    \mathbf{D}_{\text{stat},\ ii} = \sigma^2_{m_B,\, i} + \alpha'^2\, \sigma^2_{X_1,\,i} + \beta^2\,\sigma^2_{C,\,i} + C_{m_B\,X_1\,C,\,i} + \sigma^2_{\text{pec}\,i} + \sigma^2_{\text{lens},\,i} + \sigma^2_{\text{coh},\,i}\,.
\end{equation}
The matrices $\mathbf{C}_\text{stat}$ and $\mathbf{C}_\text{sys}$ can be obtained from~\cite{JLA}, which also includes the standard deviations due to the peculiar velocities $\sigma^2_{\text{pec}\,i}$, lensing $\sigma^2_{\text{lens},\,i}$, the dispersion $\sigma^2_{\text{coh},\,i}$, and the covariance among $m_B$, $X_1$ and $C$, $C_{m_B\,X_1\,C,\,i}$. Notice that $\mathbf{C}$ depends (quadratically) on the auxiliary parameters, and thereby minimising Eq.~\eqref{eq:SN_likelihood} is not fully equivalent to a least squares fitting -- even for uniform priors.

\subsection{Quasar data}
In order to use quasars as cosmological standard candles, we follow the program outlined in Refs.~\cite{Risaliti:2015zla,Risaliti:2018reu}, and which is founded on an empirical log-linear relation among the UV and X-ray luminosities,
\begin{equation}\label{eq:quasar_loglinrelation}
    \log_{10}(L_\text{X}) = \gamma\, \log_{10} (L_\text{UV}) + \text{const}\,.
\end{equation}
This translates into observable fluxes $F = L / \left[4\pi d_L(z)^2\right]$ as
\begin{equation}\label{eq:quasar_fluxes}
    \log_{10} (F_\text{X}) = \gamma\, \log_{10} (F_\text{UV}) + \beta' + 2 (\gamma-1) \underbrace{\log_{10} \left(\frac{d_L(z)}{1\Mpc}\right)}_{=\mu/5-5}.
\end{equation}
Here, $\beta'$ can in principle be related to the constant in Eq.~\eqref{eq:quasar_loglinrelation}, but an overall normalisation of $\mu$ remains undetermined~\cite{Risaliti:2015zla}. Therefore, we treat $\beta'$ as another auxiliary parameter to be fitted with the cosmology. The parameter $\gamma$ in turn can be determined from a linear fit of the flux data. In order for the redshift-dependence to be negligible, this must be carried out in narrow redshift bins, $\Delta[\log z] < 0.1$, or assuming a standard cosmology~\cite{Risaliti:2015zla}. This yields a mean of $\gamma = 0.634$ that we use throughout our statistical analysis. The data set we employ is described in Ref.~\cite{Risaliti:2018reu} and has already undergone a number of pre-selection steps, which leave a total of $N=1598$ quasar samples with redshifts $0.036 < z < 5.1$. Our likelihood function is 
\begin{equation}
    -2 \log \mathcal{L}_\text{quasar} = \sum_{i=1}^{N} \left\lbrace\frac{\left[\mu_i - \mu_\text{model}(z_i) \right]^2}{\sigma_i^2} + \log(\sigma_i) \right\rbrace,
\end{equation}
with the observed $\mu$ obtained via Eq.~\eqref{eq:quasar_fluxes} and the standard deviation $\sigma^2 = \left[\frac{5}{2\, (1-\gamma)} \, \Delta F_\text{X}\right]^2 + \delta^2$ is augmented by a dispersion parameter~$\delta$, which is included in the cosmological fit as a nuisance parameter.

\subsection{BAO data}

In the early Universe, the interaction of the relativistic photon plasma with the cooling baryons leads to density oscillations which imprint a characteristic length scale onto the CMB and also the large scale structure (LSS) of the universe. This scale can be measured as a characteristic angular scale, a \emph{standard ruler}. The rather recent measurements of the BAO scale provide an independent and complementary probe of the base cosmological model and in promise more precision and reach with upcoming surveys, such as EUCLID. Recent BAO analyses~\cite{Aubourg:2014yra,Cuceu:2019for,Leloup:2019fas,Schoneberg:2019wmt} have shown that measuring the BAO scale  is a powerful tool for probing cosmological models.

The relevant length scale for BAO is the sound horizon at the end of the so-called drag epoch $z_d$, which is the time when the photon and baryon components of the primordial plasma decouple,
\begin{equation}\label{eq:rd_def}
    r_d \equiv r_s(z_d) = \int_{z_d}^\infty \D z\, \frac{c_s(z)}{H(z)} = z_d \int_0^1 \D x \, x^{-2} \,\frac{c_s(z_d/x)}{H(z_d/x)},  
\end{equation}
where the integral expressed in terms of the variable $x=\frac{z_d}{z}$ is more suitable for numerical integration, and the sound speed given by
\begin{equation}
    c_s(z) = \frac{1}{\sqrt{3}} \left[ 1 + \frac{3}{4}\, \frac{\rho_b(z)}{\rho_\gamma(z)}\right]^{-\frac{1}{2}}.
\end{equation}
$\rho_b$ is the physical baryon density and $\rho_\gamma$ the energy density of the photon plasma. The photon density is determined from the CMB temperature $T_\text{CMB} = 2.7255$\,K~\cite{Fixsen_2009},
\begin{equation}
    \frac{3}{4 \, \Omega_\gamma h^2} = 31500 \times (T_\text{CMB} / 2.7\,\text{K})^{-4}.
\end{equation}
Notice also that at the end of the drag epoch $z_d$ the energy density of radiation in $H(z)$ cannot be ignored.

\begin{table}[h]
    \centering
    \setlength{\tabcolsep}{4pt}
    \resizebox{\textwidth}{!}{
    \begin{tabular}{r|c|ccccc|c|c}
         Name & $z_\text{eff}$ & $d_V / r_d$  & $\frac{d_M}{\Mpc} \frac{r_{d,\text{fid}}}{r_d}$   & $\frac{d_A}{\Mpc} \frac{r_{d,\text{fid}}}{r_d}$  &   $\frac{H(z) \, r_d / r_{d,\text{ fid}}}{\mathrm{km}\, \mathrm{s}^{-1}\mathrm{Mpc}^{-1}}$  & $d_H / r_d$  &      $\frac{r_{d,\text{ fid}}}{\mathrm{Mpc}}$ &   $r_\text{corr}$\\ 
         \midrule
         6dFGS~\cite{Beutler:2011hx}  &   0.106   &   $2.976 \pm 0.133$   &   $-$ &   $-$    &   $-$ &   $-$ &   $-$    &   $-$\\
         SDSS MGS~\cite{Ross:2014qpa}    &   0.15    &   $4.466 \pm 0.168$   &   $-$ &   $-$ &   $-$    &   $-$ &   $148.69$    &   $-$\\
         \midrule
            &   0.38    &   $-$   &  $1518 \pm 22$  &   $-$ &  $81.5 \pm 1.9$     &   $-$   &  $147.78$   &   \multirow{3}{*}{\rotatebox{90}{cov.~matrix}}\\
         BOSS DR12~\cite{Alam:2016hwk} &   0.51    &   $-$ &  $1977 \pm 27$   &   $-$ &  $90.4 \pm 1.9$     &   $-$   &  $147.78$ & \\
            &   0.61    &   $-$ &  $2283 \pm 32$  &   $-$ &  $97.3 \pm 2.1$     &   $-$   &  $147.78$   &    \\
         \midrule
         BOSS DR14~\cite{Bautista:2017wwp}  &   0.72    &   $16.08 \pm 0.41$    &  $-$  &    $-$  &  $-$  &   $-$    &    $147.78$    &   $-$\\
         \midrule
         \multirow{4}{*}{eBOSS QSO~\cite{Zhao:2018jxv}}   &   0.978    &   $-$ &  $-$ &  $1586 \pm 284$ &     $113.72 \pm 14.63$ &   $-$  &   $147.78$ &  \multirow{4}{*}{\rotatebox{90}{cov.~matrix}}\\
            &   1.23    &   $-$ &   $-$  &   $1769 \pm 160$  &   $131.44 \pm 12.42$ &   $-$  &   $147.78$ &   \\
            &   1.526    &   $-$ & $-$  &   $1768.8 \pm 96.6$  &   $148.11 \pm 12.75$ &   $-$  &   $147.78$ &   \\
            &   1.944    &   $-$ &   $-$ &   $1808 \pm 146$  &   $172.63 \pm 14.79$ &   $-$  &   $147.78$ &   \\
         \midrule
         eBOSS Ly$\alpha$~\cite{Blomqvist:2019rah}    &   2.34    &   $-$ &   $(37.41 \pm 1.86) \, r_{d,\text{fid}}$& $-$  &   $-$    &   $8.86 \pm 0.29$  &   $147.33$   &   $-0.34$\\
         eBOSS Ly$\alpha$-QSO~\cite{Agathe:2019vsu}   &   2.35    &   $-$ &   $(36.3 \pm 1.8)\,  r_{d,\text{fid}}$&   $-$    &   $-$ &   $9.20 \pm 0.36$  &   $147.33$   &   $-0.44$\\
         eBOSS Ly$\alpha$ combined~\cite{Agathe:2019vsu}   &   2.34    &   $-$ &   $(37.1 \pm 1.2) \, r_{d,\text{fid}}$&   $-$    &   $-$ &   $9.00 \pm 0.22$  &   $147.33$   &   $-0.40$\\
         \bottomrule
    \end{tabular}}
    \caption{BAO measurements used in our analysis. This table is adapted from Ref.~\cite{Aubourg:2014yra} with updated data sets as found in Ref.~\cite{Cuceu:2019for}. The correlation matrices can be found in the references.}
    \label{tab:BAOdata}
\end{table}



The dynamics of the drag epoch have been thoroughly analysed in~\cite{Hu:1995en}, where a numerical fitting formula for $z_d$ is given,
\begin{align}
    z_d & = 1345 \,\frac{(\Omega_m h^2)^{0.251}}{1 + 0.659 \, (\Omega_m h^2)^{0.828} }\,[1 + b_1 (\Omega_b h^2)^{b_2}], \nonumber\\
 b_1  &= 0.313 \,(\Omega_m h^2)^{-0.419} \,[1 + 0.607 (\Omega_m h^2)^{0.674} ], \nonumber\\
 b_2 &= 0.238 \,(\Omega_m h^2)^{0.223}.
\end{align}

The relevant cosmological distance measure for an object of known size is the redshift-weighted comoving distance $d_M$,
\begin{equation}
    d_M(z) = (1+z) d_A(z) = \frac{d_L(z)}{1+z}.
\end{equation}
In order to measure the BAO scale, a fiducial cosmology is employed that allows to translate the power spectrum to a distance measure, while allowing the BAO scale to shift relative to the fiducial cosmology's prediction. The measurements are then typically quantified by one or two numbers, that quantify the discrepancy between the measured BAO angle and the fiducial BAO angle. In the case of an anisotropic survey, this yields a measurement perpendicular to the line of sight and one parallel to it:
\begin{equation}
    \frac{d_M(z)}{r_d} = \alpha_\perp \frac{d_{M,\ \text{fid}}(z)}{r_{d,\ \text{fid}}}, \qquad \text{and} \qquad
    \frac{d_H(z)}{r_d} = \alpha_\parallel \frac{d_{H,\ \text{fid}}(z)}{r_{d,\ \text{fid}}},
\end{equation}
with $d_H(z) = c/H(z)$, while isotropic surveys constrain a single quantity defined as
\begin{equation}
    \frac{d_V(z)}{r_d} = \alpha \frac{d_{V,\ \text{fid}}(z)}{r_{d,\ \text{fid}}},
\end{equation}
with $d_V(z) = [z\, d_H(z)\, d_M^2(z)]^\frac{1}{3}$ a volume averaged distance measure. In Tab.~\ref{tab:BAOdata} we present all measurements that have been taken into account in our study.

In summary, the BAO likelihood piece is
\begin{equation}
    -2 \log \mathcal{L}_\text{BAO} = \left[ \vec{Y} - \vec{Y}_\text{model} \right]^T \mathbf{C}^{-1}_\text{BAO}\left[ \vec{Y} - \vec{Y}_\text{model} \right],
\end{equation}
where $\vec{Y}$ is a vector containing the measured quantities in Tab.~\ref{tab:BAOdata} and $\mathbf{C}_\text{BAO}$ is a matrix of correlations assembled also from it.


\paragraph{CMB anisotropies as BAO measurement.} Finally, we treat the measurement of the first peak in the CMB spectrum as a BAO experiment at redshift $z_*$. This is a well-established procedure, which was also used in the analysis of SN data in Ref.~\cite{JLA}, BAO data in Ref.~\cite{Aubourg:2014yra}.
For our purposes, we use the Planck 2018 results.

The redshift of last scattering is approximated as in~\cite{Hu:1995en} by
\begin{align}
z_* & = 1048 \, [1 + 0.00124 (\Omega_b h^2)^{-0.738}]\,[1 + g_1 \,(\Omega_m h^2)^{g_2} ] , \nonumber \\
g_1 & = 0.0783 \, (\Omega_b h^2)^{-0.238}\, [1+39.5\,(\Omega_b h^2)^{0.763}]^{-1} , \nonumber \\
g_2 & = 0.560 \, [1 + 21.1\, (\Omega_b h^2)^{1.81}]^{-1} .
\end{align}
Crucially, the redshift $z_d$ which sets the end of the drag epoch and that of last scattering~$z_*$ are not exactly equal, with $z_* \gtrsim z_d$. For example, we find $z_* = 1092$ and $z_d = 1063$ for the best fit \LCDM~cosmology. This affects the comoving sound horizon at the percent level.

The CMB data is implemented in the form of distance priors, which compress the information of the full parameter chains inferred from the final Planck~2018 data. They have been calculated in Ref.~\cite{Chen:2018dbv} for the cross correlation of TT, EE, TE + lowE power spectra. For the base model \LCDM, this is
\begin{equation}
    \vec{X}^T \equiv \left(R , l_A, \Omega_b h^2\right) = (1.7502, 301.471, 0.02236)\,,
\end{equation}
and a marginalised inverse correlation matrix is obtained 
\begin{equation}
     \mathbf{C}_\text{Planck} = 10^{-5}\cdot \left(
    \begin{tabular}{ccc}
        $2.1$ &  $19$ & $-0.045$ \\
       $ 19$ & $789$ &  $-0.43$, \\
       $-0.045$ &  $-0.43$ & $0.0022$ 
    \end{tabular}\right).
\end{equation}
See \cite{Chen:2018dbv} for the distance priors for cosmologies including curvature and dynamical dark energy, which we have implemented as well. The parameters $R$ and $l_A$ are determined by the cosmology as
\begin{align}
    R & = \frac{H_0}{c} \sqrt{\Omega_m} \, (1+z_*) \, D_A(z_*) \nonumber\\
    l_A & = (1+z_*) \frac{\pi D_A(z_*) }{r_s (z_*)}.
\end{align}
Thus, for all analyses labelled  `+CMB', we include in the likelihood function a factor
\begin{equation}
     -2 \log \mathcal{L}_\text{CMB} = \left[ \vec{X} - \vec{X}_\text{model} \right]^T \mathbf{C}^{-1}_\text{Planck} \left[ \vec{X} - \vec{X}_\text{model} \right]
\end{equation}

\subsection{Joint analysis of cosmological data}
To combine the SN and quasar likelihoods, we assume the data to be independent and thus multiply the probabilities, or equivalently add the log-likelihoods,
\begin{equation}
   \log \mathcal{L}_\text{tot} = \log \mathcal{L}_\text{SN} + \log \mathcal{L}_\text{quasar} + \log \mathcal{L}_\text{BAO} + \log \mathcal{L}_\text{CMB}\,.
\end{equation}
We sample the posterior probability distribution,
\begin{equation}
    p(\theta|\vec{x}) \propto p(\theta) \, \mathcal{L}(\vec{x}|\theta)
\end{equation}
assuming uniform prior distributions $p(\theta)$ for the auxiliary as well as the  cosmological parameters $\theta = (\alpha',\beta,M_B^0,\Delta M_B,\beta',\delta,\Omega_m,\Omega_\Lambda,w,\ldots)^T$. To this end, we make use of the python package \texttt{emcee}, which implements an affine invariant MCMC ensemble sampler~\cite{ForemanMackey:2012ig}, a technique particularly well suited for our purposes~\cite{Goodman}. 

In order to quantitatively compare models, we employ the so-called `Bayes information criterion' (BIC), which takes into account not only how well a model fits a data but also its simplicity in terms of the number of parameters it introduces:
\begin{equation}
    \mathrm{BIC} \equiv |\theta| \log (|\vec{x}|) - 2\,\log(\hat{\mathcal{L}}),
\end{equation}
where $\hat{\mathcal{L}}$ is the maximised value of the posterior probability distribution and $|\cdot|$ denotes the length of a vector. In order to select among two models the preferred one, we compare the \emph{evidence} of the data occurring within a given model, 
\begin{equation}\label{eq:BICdef}
    p(\vec{x}|M) \equiv \int \D\theta\, {p(\theta_M)\, p(\vec{x}|\theta_M)} \,,
\end{equation}
where $\theta_M$ represents the vector of parameters in a given model $M$. It is in general not possible to directly evaluate this integral, so either this has to done via another MCMC approach, or an approximate method. It can be shown that the BIC, as defined in Eq.~\eqref{eq:BICdef}, serves as an estimator of the evidence
\begin{equation}\label{eq:BICmeaning}
     p(\vec{x}|M) \propto \exp(-\mathrm{BIC}/2)\,.
\end{equation}
To see this, we expand the posterior into a Taylor series to second order about the point of maximal likelihood, i.e.
\begin{equation}
    \log p(\theta|\vec{x}) \propto \log\left[ p(\theta) \, \mathcal{L}(\vec{x}|\theta)\right] \approx \log p(\hat{\theta}) \, \mathcal{L}(\vec{x}|\hat{\theta}) - \frac{1}{2} (\theta - \hat{\theta})_i\, \tilde{H}_{ij} \,(\theta - \hat{\theta})_i\,,
\end{equation}
with $\tilde{H}$ the negative Hessian of the posterior $p(\theta|\vec{x})$ evaluated at the parameter value $\theta = \hat{\theta}$ that maximises it.\footnote{The first order term vanishes due to the maximum condition, and we define $\tilde{H} = - H$, such that it is positive definite.}
The integral in Eq.~\eqref{eq:BICdef} is now a multidimensional Gaussian and yields
\begin{equation}
    \log p(\vec{x}|M) = \log p(\hat{\theta}) + \log \mathcal{L}(\vec{x}|\hat{\theta}) +\frac{1}{2} \log \left( \frac{2\,\pi^{|\hat{\theta}|}}{\det \tilde{H}} \right).
\end{equation}
Note that $\tilde{H}$ is the Fisher information matrix, which one can show factorizes as $\tilde{H} = n \tilde{I}$, where $\tilde{I}$ is the Fisher information matrix for a single data point~\cite{BIC}. Thus, in the limit of large $n$,
\begin{equation}
    \log p(\vec{x}|M) = \log p(\hat{\theta}) + \log \mathcal{L}(\vec{x}|\hat{\theta}) +\frac{1}{2} \log \left( \frac{2\,\pi^{|\hat{\theta}|}}{n^{|\hat{\theta}|}\det \tilde{I}} \right).
\end{equation}
Taking the asymptotic limit $n\to\infty$ and ignoring all terms that do not scale with $n$, one finds,
\begin{equation}
    -2\,\log p(\vec{x}|M) = |\hat{\theta}| \log(n) - 2\,\log \mathcal{L}(\vec{x}|\hat{\theta})\,,
\end{equation}
from which Eq.~\eqref{eq:BICmeaning} emerges. Therefore, the probability of erroneously choosing Model $M$ over model $M'$ can be estimated as
\begin{equation}
    P(M) = \frac{e^{-\mathrm{BIC}(M)/2}}{e^{-\mathrm{BIC}(M)/2} + e^{-\mathrm{BIC}(M')/2}} = \frac{1}{1 + e^{-\Delta/2}}\,, \quad \text{with $\Delta = \mathrm{BIC}(M) - \mathrm{BIC}(M')$}.
\end{equation}
Thus, if $\Delta =1.5/5.9/11.6$ there is mild/strong/very strong evidence to reject model $M$ in favour of model $M'$, corresponding to $1-P(M) = 68\%/95\%/99.7\%$~CL, respectively. 

\bibliographystyle{JHEP}
\bibliography{literature}

\end{document}